\pdfoutput=1
\documentclass[12pt]{iopart}

\expandafter\let\csname equation*\endcsname\relax
\expandafter\let\csname endequation*\endcsname\relax

\usepackage{graphicx}
\usepackage{siunitx}
\usepackage{physics}
\usepackage[hidelinks]{hyperref}
\usepackage{makecell}

\usepackage[acronym, nomain]{glossaries}
\newacronym{rf}{RF}{radio frequency}
\newacronym{nisq}{NISQ}{Noisy Intermediate-Scale Quantum}
\newacronym{QCCD}{QCCD}{quantum charge-coupled device}
\newacronym{se}{SE}{Surface-Electrode}
\newacronym{fem}{FEM}{Finite Element Method}
\newacronym{tsv}{TSV}{Through-Substrate Via}
\newacronym{emm}{EMM}{excess micromotion}
\newacronym{roi}{ROI}{region of interest}
\newacronym{dac}{DAC}{Digital-Analog Converter}

\newcommand{\mm}[1]{\SI{#1}{\milli\meter}}
\newcommand{\um}[1]{\SI{#1}{\micro\meter}}
\newcommand{\nm}[1]{\SI{#1}{\nano\meter}}
\newcommand{\kHz}[1]{\SI{#1}{\kilo\hertz}}
\newcommand{\MHz}[1]{\SI{#1}{\mega\hertz}}
\newcommand*{\SiN}{Si\textsubscript{3}N\textsubscript{4}}

\begin{document}

\title{Optimization and implementation of a surface-electrode ion trap junction}

\author{Chi Zhang, Karan K Mehta and Jonathan P Home}

\address{Institute for Quantum Electronics, ETH Z\"urich, Otto-Stern-Weg 1, 8093 Z\"urich, Switzerland}
\ead{\mailto{zhangc@phys.ethz.ch} and \mailto{jhome@phys.ethz.ch}}
\vspace{10pt}
\begin{indented}
\item[] January 2022
\end{indented}

\begin{abstract}
We describe the design of a surface-electrode ion trap junction, 
which is a key element for large-scale ion trap arrays.
A bi-objective optimization method is used for designing the electrodes, which maintains the total pseudo-potential curvature while minimizing the axial pseudo-potential gradient along the ion transport path.
To facilitate the laser beam delivery for parallel operations in multiple trap zones, we implemented integrated optics on each arm of this X-junction trap. 
The layout of the trap chip for commercial foundry fabrication is presented.
This work suggests routes to improving ion trap junction performance in scalable implementations.
Together with integrated optical addressing, this contributes to modular trapped-ion quantum computing in interconnected 2-dimensional arrays.
\end{abstract}

%
\vspace{2pc}

\section{Introduction}
Trapped ions are a leading platform for quantum computation.
All the elementary building blocks for a quantum computer have been realized in these systems with high quality, including efficient and robust state preparation and readout~\cite{Harty2014}, high-fidelity universal quantum gates~\cite{Clark2021,Srinivas2021} and long qubit memory times~\cite{Wang2017,Sepiol2019}. 
This has allowed the demonstration of small-scale computation tasks~\cite{Debnath2016,Monz2016,Nam2020},
as well as elements of fault-tolerant control~\cite{Egan2021,postler2021demonstration} and quantum error correction~\cite{Negnevitsky2018,RyanAnderson2021}, which establish reliable primitives for scaling.

To build a trapped-ion quantum computer capable of outperforming its classical counterpart in practically useful applications,
a much larger number of ions are required~\cite{Preskill1998}.
Scaling up the number of ions in a single linear trap faces several challenges,
such as difficulty in individual ion addressing due to the reduced inter-ion spacing~\cite{James1998Quantum},
slowing down of multi-qubit gates~\cite{Murali2020},
and crowding of motional modes~\cite{wineland1998experimental} that can reduce the fidelity of quantum logic operations due to the thermal occupation of spectator modes~\cite{wineland1998experimental,Wu2018} or cross-Kerr coupling between modes~\cite{Nie2009}.
These challenges necessitate interconnected ion-trap modules, each hosting a relatively small number of ions,
to maintain the low error rates when scaling up. 

One way of interconnecting elementary trapped-ion logic units is via photonic links~\cite{Monroe2014, Hucul2015,Stephenson2020},
which is likely to be a necessary step for establishing a truly large-scale trapped-ion quantum computer.
The \gls{QCCD} architecture~\cite{Kielpinski2002} provides an intermediate level of trap interconnection that is promising for controlling $>100$ trapped-ion qubits,
where the interconnection among elementary logic units is realized by ion shuttling.
Basic quantum error correction and quantum algorithms can be realized on \gls{QCCD} nodes, 
and photonics links could then be used to mediate communication between several modules~\cite{Nickerson2014,Nigmatullin2016}.

Small-scale qubit control based on the \gls{QCCD} architecture have been demonstrated on linear traps with up to 20 ions~\cite{RyanAnderson2021,Home2009,Kaufmann2017,Pino2021}. 
In the largest such demonstrations, swap operations were used to change the order of the ion chain and bring two arbitrary ions into the interaction zone~\cite{RyanAnderson2021}.
In a larger system, such a 1-dimensional \gls{QCCD} architecture demands a large number of swap operations to interact two arbitrary ions, 
especially when there are intermediate ion crystals between them.  
For applications with more irregular communication patterns,
a 2-dimensional grid of ion traps interconnected by junctions is likely to provide significant performance benefits~\cite{Murali2020}.

Junction traps have been demonstrated, for instance 
transportation with 0.05 to 0.1 quanta of average motional excitation per junction traverse has been realized using a 3-dimensional trap fabricated from a stack of laser-machined alumina wafers \cite{Blakestad2009,Blakestad2011}. 
More advanced laser machining technologies~\cite{Ragg2019} have been used to implement a 3-dimensional wafer trap with double junctions~\cite{decaroli2021design}.
However, the complexities in assembling these 3-dimensional wafer traps hamper their scalable reproduction.
By contrast, lithographic methods allow asymmetric \gls{se} traps~\cite{Wesenberg2008},
where all trap electrodes are patterned on a single plane,
to leverage advanced and scalable semiconductor manufacturing technologies~\cite{Pino2021,Mehta2014,SandiaJunction2011}.
Several \gls{se} junction traps have been designed and fabricated~\cite{SandiaJunction2011,Amini2010,Wright2013GTRI_junction},
but by far the best reported performance in junction transportation is 37 to 150 quanta of motional excitation per junction traverse~\cite{Shu2014}.
This is much higher than the $\sim 0.1$ quanta level achieved in the 3-dimensional trap~\cite{Blakestad2011} and limits the realization of \gls{QCCD} architecture based on \gls{se} junction traps, pointing to the need for further research in this area.

In the direction perpendicular to the trap surface, 
\gls{se} traps lack the mirror or rotation symmetry compared to their 3-dimensional counterparts. 
This naturally reduces the trap depth~\cite{Wesenberg2008} 
and induces anharmonicity at the ion position~\cite{Home2011},
imposing more challenges in ion transport.
Despite these challenges, ion transport with sub-quantum excitation has been realized on the linear section of the \gls{se} junction trap~\cite{Shu2014} or linear \gls{se} traps~\cite{Pino2021}.
This suggests the sub-optimal junction transport on \gls{se} traps  is limited by the junction design, 
rather than the fundamental trap symmetry.

Early \gls{se} junction designs mainly emphasized the minimization of the pseudo-potential barrier~\cite{SandiaJunction2011,Amini2010,Wright2013GTRI_junction},
an inevitable feature near the junction center for a \gls{rf} quadrupole trap (see \sref{sec:junction_principle}).
Consequently, the total pseudo-potential curvature is also reduced near the junction center.
This enforces a reduction of the secular frequencies or the
lowering of the ion height from the initially optimized transport path~\cite{Wright2013GTRI_junction},
and makes the trap operations more susceptible to stray electric fields~\cite{Amini2010,Wright2013GTRI_junction}.
In the designs presented here, we decided to follow an alternative approach,
placing more weights on maintaining sufficient pseudo-potential curvature during the optimization of the \gls{rf} electrodes close to the junction. 
In this paper, we describe the design process for this trap, considering also the use of integrated optics on each arm of the X-junction trap to facilitate laser beam delivery, a feature which will be of increasing importance as the number of trapping zones increases. The designs have been recently fabricated in a commercial photonic foundry.

The rest of this paper is organized as follows.
\Sref{sec:junction_RF} presents the design principle of the \gls{rf} electrodes, the optimization procedures, and the verification of the design.
\Sref{sec:junction_control} describes the design of the control electrodes.
In \Sref{sec:integrated_optics}, we present the integrated optics implemented on the junction traps.
The full trap chip layout is presented in \sref{sec:chip_layout}. 
In \sref{sec:summary}, we conclude with an outlook for experiments with such junction traps and other technologies that can potentially benefit the future iterations.

\section{RF electrodes of junction trap}
\label{sec:junction_RF}
Since the quasi-static trapping potential can typically be flexibly adjusted during the experiment, 
whereas the spatial \gls{rf} potential is fixed by the electrode geometry for a single non-adjustable \gls{rf} drive,
in the design process of the junction trap we mainly focused on the \gls{rf} electrodes.
Once this design stage is complete, we then focus our attention on optimizing the patterning of the control electrodes,
to verify that the desired total trapping potential can be created with static voltages feasible for typical \glspl{dac} used in experiments today~\cite{Creotech,Clercq2015}.

\subsection{Design principle}
\label{sec:junction_principle}

To achieve zero electric field everywhere along two intersecting ion transport paths that are confining,
a hexapole is the lowest electric multipole term allowed at the intersection~\cite{Wesenberg2009}.
Unfortunately, an ion at the center of a pure hexapole potential has zero harmonic motional frequency~\cite{hexapole2006}.
If a junction trap creates non-zero \gls{rf} quadrupole confinement at the intersection, 
there is inevitably non-vanishing \gls{rf} field along any ion path in vicinity of the intersection. 
Due to the broken axial translational symmetry near the intersection, 
this non-vanishing \gls{rf} field is inhomogeneous along the trap axes,
giving rise to the \textit{pseudo-potential barrier}~\cite{Blakestad2009},
with the pseudo-potential given as~\cite{Dehmelt1968}
\begin{eqnarray}
    \phi_\textrm{PP} =\frac{Q}{4m\Omega_\text{RF}^2}\abs{\vec{E}_\textrm{RF}}^2
\end{eqnarray}
for an ion of mass $m$ and charge $Q$ within an \gls{rf} field $\vec{E}_\textrm{RF}$ of angular frequency $\Omega_\textrm{RF}$.
For several reasons, it is beneficial to reduce the height of the pseudo-potential barrier.
A high $\phi_\textrm{PP}$ causes a large amplitude of the driven motion of the ion (typically referred to as \textit{micromotion}), which could break down the pseudo-potential approximation~\cite{gerlich1992inhomogeneousField}.  
The gradient of $\phi_\textrm{PP}$ associated with the pseudo-potential barrier can cause heating of the secular motional modes when noise on the \gls{rf} potential is present~\cite{Blakestad2009,Wright2013GTRI_junction,Mokhberi2017}.
Due to the mass dependence of $\phi_\textrm{PP}$, the pseudo-potential barrier may also cause segregation of ions of different species when transporting a mixed-species crystal across the junction~\cite{Mokhberi2017}.

Despite the pseudo-potential barrier,
an \gls{rf} quadrupole trap has advantages over higher-order traps in offering a higher overall curvature and intrinsic trap depth~\cite{Wesenberg2008}.
Since the pseudo-potential barrier has been demonstrated not to be a fundamental limit of high-quality ion transport across junction~\cite{Blakestad2009,Blakestad2011},
we have taken the approach of designing an \gls{rf} quadrupole trap.

The pseudo-potential curvature is one important factor often overlooked in junction designs.
The total quasi-static potential can be written as the superposition of the pseudo-potential $\phi_\textrm{PP}$ and the quasi-static potential $\phi_\text{st}$ from the control electrodes
\begin{eqnarray}
    \phi_\text{total}(\vec{r}) = \phi_\text{PP}(\vec{r}) + \phi_\text{st}(\vec{r}).
\end{eqnarray}
Because $\phi_\text{st}$ satisfies the Laplace equation, the Laplacian of $\phi_\text{total}$ follows
\begin{eqnarray}
    \laplacian\phi_\text{total}(\vec{r})= \laplacian\phi_\text{PP}(\vec{r}),
    \label{def:total_confinement}
\end{eqnarray}
which suggests that the overall positive curvature for confining the ions in three dimensions is created solely by the pseudo-potential. 
The quasi-static potential $\phi_\text{st}$ can redistribute the total pseudo-potential curvature among the three directions,
but does not affect the net confinement of the trap.
If the pseudo-potential confinement is sacrificed in the minimization of the pseudo-potential barrier,
the secular frequencies must be reduced close to the junction center.
The ions will thus be more susceptible to the drift of stray field~\cite{Amini2010} and is likely to experience higher static motional heating due to electric noise~\cite{Brownnutt2015Heating}.

In the numerical optimization of the junction \gls{rf} electrodes, we therefore seek for a balanced performance in the pseudo-potential barrier height and the total confinement.
Our numerical optimization is based on the \textit{SurfacePattern} software package~\cite{Schmied2009,Schmied2010SurfacePattern},
which solves the electrostatic potential of planar electrodes assuming 
\begin{enumerate}
    \item that there is no gap between electrodes;
    \item that the entire plane is covered by conducting electrodes.
\end{enumerate}
As will be seen in \sref{sec:junction_comsol}, the positions of pseudo-potential minima obtained with \textit{SurfacePattern} model agrees to within 5\% with the \Gls{fem} model of the trap, 
and the worst-case discrepancy in pseudo-potential is about 20\% around the junction center.
These discrepancies are well understood to be due to the \um{5} gaps between the trap electrodes in our fabrication technology.
We chose to use the \textit{SurfacePattern} model at the initial stage of the design process, because its analytical expressions of the electric potentials allow for faster evaluations and iterations of the design. 
The more precise \gls{fem} model is used to verify the optimized \gls{rf} electrode geometry.

\subsection{Parameterization of electrode geometry}
\label{sec:junction_spline}
To establish a parametric description of the electrode geometry, we first took the \textit{control point} approach similar to \cite{Amini2010,Wright2013GTRI_junction,Mokhberi2017,GTRI2012},
where the electrode boundaries are defined by straight line segments connecting neighbouring control points.
We noticed two limitations with this approach:
(1) Sharp corners were formed at the control points, which will likely create high \gls{rf} field emission that reduces the breakdown voltage of the trap \cite{JacksonClassicalElectrodynamics},
although these structures with high spatial frequency $k_\text{s}$ will not significantly affect the potential landscape at the ion position $r\gg(2\pi)/k_\text{s}$,
as their contributions exponentially decrease as $\exp(-k_\text{s}r)$~\cite{Wesenberg2008}. 
(2) To avoid creating self-intersecting boundaries, the allowed range for each control point had to be chosen with care, which made it difficult to ensure the obtained geometry is close to a global optimum in the full parameter space. 

\begin{figure}[htb]
\centering
\includegraphics[width=2.5in]{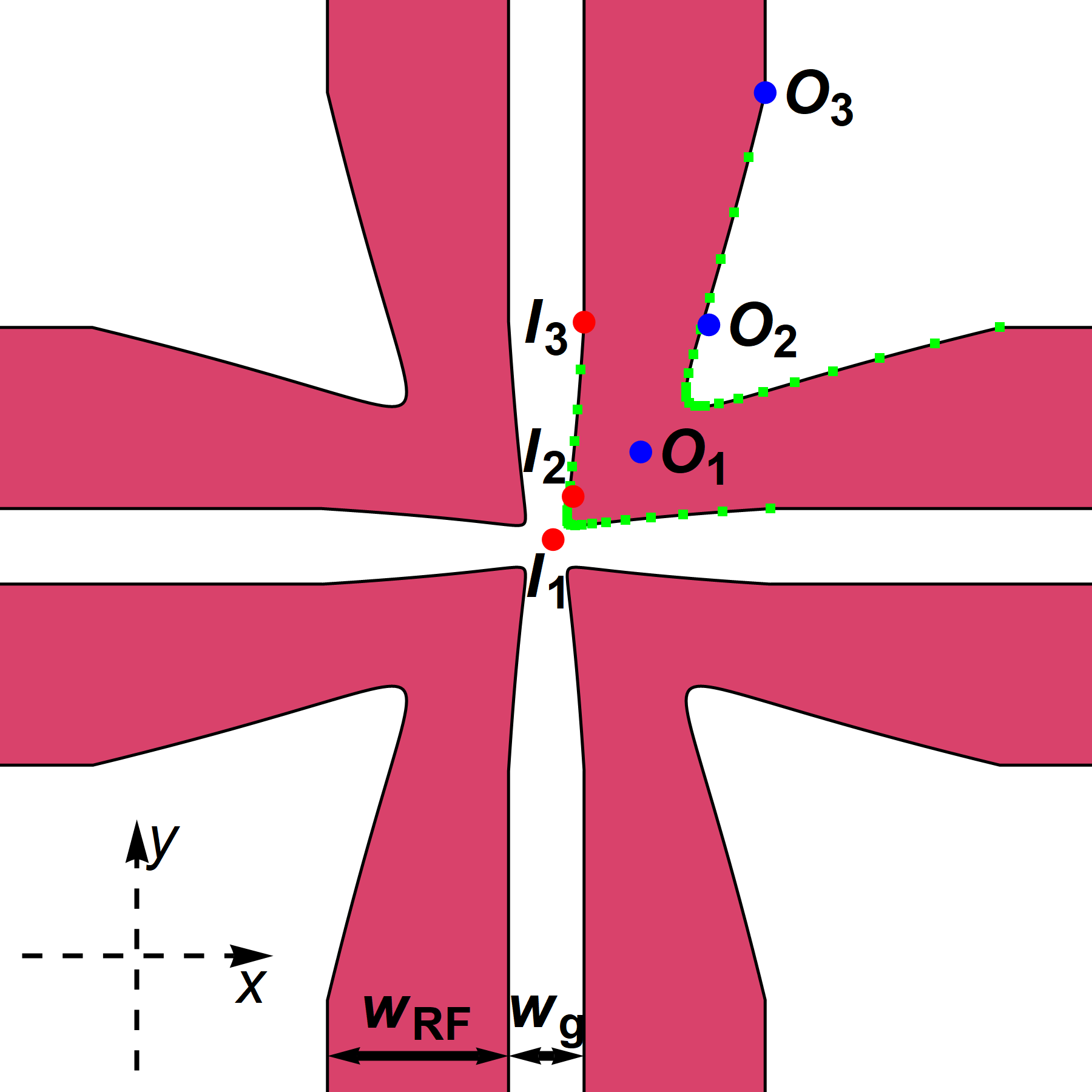}
\caption{Spline function parameterization of the X-junction.
The colored areas indicate the \gls{rf} electrode.
The rest of the surface is modeled to be  \gls{rf} grounded. 
The independent control points of the inner (outer) spline are indicated with red (blue) dots.
The 21 sample points of each spline are indicated in green.
The independent variables of the control points are listed in \tref{tab:junction_spline}.
The origin of the coordinate system is set at the center of the junction.
$z$ axis is defined upwards perpendicular to the trap surface.
}
\label{fig:junction_spline}
\end{figure}

A parameterization based on spline functions~\cite{Mokhberi2017} solves these problems.
Our optimized geometry using the spline parameterization is shown in \fref{fig:junction_spline}.
The inner and outer boundaries of the \gls{rf} electrode are respectively described by a B-spline function of cubic order.
Each spline function is defined by 5 control points, 
with reflection symmetry about the diagonal line $y=x$ in \fref{fig:junction_spline}.
Hence only the 3 independent control points are shown for each spline.
The coordinates of these control points and their independent variables are listed in \tref{tab:junction_spline}.
During the numerical optimization, we sample each spline function with 21 points evenly distributed in the spline parameter space (green points in \fref{fig:junction_spline}),
and define the \gls{rf} electrode as a polygon by connecting the neighbouring sample points with a straight line segment.

\begin{table}[ht]
\centering
\caption{Control points of spline function parameterization in \fref{fig:junction_spline} and the independent variables.} 
\label{tab:junction_spline}
\begin{indented}
\item[]\begin{tabular}{@{}lll@{}} 
\br
Point & Coordinate & Independent variable \\ 
\mr
$I_1$ & $(d_\text{in},d_\text{in})$ & $d_\text{in}$\\
$I_2$ & $\qty(x_\text{i2},y_\text{i2})$ & $x_\text{i2}$, $y_\text{i2}$\\
$I_3$ & $\qty(w_\text{g}/2, y_\text{i3})$ & $y_\text{i3}$\\
$O_1$ & $(d_\text{out},d_\text{out})$ & $d_\text{out}$\\
$O_2$ & $\qty(x_\text{o2},y_\text{o2})$ & $x_\text{o2}$, $y_\text{o2}$\\
$O_3$ & $\qty(w_\text{g}/2+w_\text{RF}, y_\text{o3})$ & $y_\text{o3}$\\
\br
\end{tabular}
\end{indented}
\end{table}

We chose the target ion height to be $h=\um{50}$.
Using the analytical expressions for the trapping potential assuming gapless, infinitely extended surface electrodes~\cite{House2008},
we optimized the geometry of \gls{rf} electrodes on the linear sections for maximal pseudo-potential curvature at fixed height $h$.
The resulting width ($w_\textrm{RF}$) and separation ($w_\textrm{g}$) of the \gls{rf} electrodes are listed in \tref{tab:junction_linear},
which agree with \cite{Mokhberi2017} to within 99.5\%
with the slight discrepancy due to numerical truncation errors in the intermediate calculation steps. 

\begin{table}[htbp]
\caption{Optimal \gls{rf} geometry on the linear arm for maximal pseudo-potential confinement at fixed ion height $h$.} 
\label{tab:junction_linear}
\begin{indented}
\item[]\begin{tabular}{@{}lll@{}}
\br
& Expression& \makecell{Value for $h=\um{50}$ \\(\si{\micro\meter})}\\ 
\mr
$w_\textrm{g}$ & $0.83h$ & 41.5\\
$w_\textrm{RF}$ & $4h^2-w_\textrm{g}^2/(2w_\textrm{g})\approx 1.99h$ & 99.5\\
\br
\end{tabular}
\end{indented}
\end{table}

\subsection{Numerical optimization}
For the parameterized junction \gls{rf} geometry, 
we performed a bi-objective optimization using the \textit{Nelder-Mead} algorithm.
The cost function is evaluated along the horizontal trap axis in \fref{fig:junction_spline} denoted by $x$, at the target trap height $h$.
One objective is to minimize the variance of the pseudo-potential confinement along the arm
\begin{eqnarray}
  f_1 = \text{Var}\qty(\laplacian\phi_\textrm{PP}(\vec{r})).
  \label{cost:pp_curvature_variance}
\end{eqnarray}
The second objective is to minimize the axial gradient of the pseudo-potential 
\begin{eqnarray}
    f_2 = \int_0^{x_\text{max}} \abs{\pdv{\phi_\textrm{PP}(x,0,h)}{x}}\dd x,
    \label{cost:axial_field}
\end{eqnarray}
which can cause heating of the axial motional mode in the presence of \gls{rf} noise.
This second objective also serves as a proxy for reducing the pseudo-potential barrier along the trap axis,
but is more efficient in evaluation,
which saves the need to find the peak of the pseudo-potential barrier first.
Similar to \cite{Mokhberi2017}, we approximated \eref{cost:axial_field} as a summation over 101 evaluation points evenly distributed along the trap axis with a spacing of $\dd x = 0.1h=\um{5}$, from the junction center $x=0$ to $x_\text{max}=10h=\um{500}$.
We used these evaluation points to calculate the variance of the pseudo-potential confinement in \eref{cost:pp_curvature_variance} as well.

In the numerical optimization, we used dimensionless length normalized with $h$,
and a dimensionless pseudo-potential defined as~\cite{Mokhberi2017} 
\begin{eqnarray}
    \phi_\textrm{PP}'=\frac{4m\Omega_\textrm{RF}^2 h^2}{Q^2 V_\textrm{RF}^2}\phi_\textrm{PP}
    =\abs{\nabla\Theta_\textrm{RF}}^2,
\end{eqnarray}
where $\Theta_\textrm{RF}$ is the basis function for the \gls{rf} electrodes~\cite{Hucul2008} that solely depends on their geometry.
The overall cost function was a weighted sum of the two contributions 
\begin{eqnarray}
   \fl f_\text{cost} = w_1 f_1 + w_2 f_2 =w_1 \text{Var}\qty(\laplacian\phi_\textrm{PP}'(x,0,h))
    + w_2 \sum_{i=0}^{100} \abs{\pdv{\phi_\textrm{PP}'(x_i,0,h)}{x}}\dd x,
    \label{eqn:dual_cost}
\end{eqnarray}
with the weights $w_1$ and $w_2$.
The final junction geometry shown in \fref{fig:junction_spline} was optimized for balanced weights 
$w_1=w_2$.

To reduce the chance of getting trapped in local minima of the parameter space,
we ran the same numerical optimization with 16 different random seeds, 
which determine the starting points in the parameter space,
and selected the outcome with the lowest cost as the final result.
Qualitatively, we also verified that other seeds resulted in geometries similar to the chosen one,
indicating a better chance that our chosen solution approaches the global optimum.
The values of the independent variables defined in \tref{tab:junction_spline} for the final geometry are listed in \tref{tab:junction_final}.

\begin{table}[htp]
\caption{Specification of the independent variables in \tref{tab:junction_spline}, for the final junction geometry put into fabrication as shown in \fref{fig:junction_spline}.} 
\label{tab:junction_final}
\begin{indented}
\lineup
\item[]\begin{tabular}{@{}lll@{}} 
\br
Variable & Expression & \makecell{Value for $h=\um{50}$\\(\si{\micro\meter})}\\ 
\mr
$d_\text{in}$ & $0.07460h$ &$\0\03.73$\cr
$x_\text{i2}$ & $0.29428h$ &\014.71\cr
$y_\text{i2}$ & $0.54857h$ &\027.43\cr
$y_\text{i3}$ & $2.46382h$ &123.19\cr
$d_\text{out}$ & $1.03774h$ &\051.89\cr
$x_\text{o2}$ & $1.78611h$ &\089.31\cr
$y_\text{o2}$ & $2.43434h$ &121.72\cr
$y_\text{o3}$ & $4.98629h$ &249.32\cr
\br
\end{tabular}
\end{indented}
\end{table}

\subsection{Verification of the final RF geometry}
\label{sec:junction_comsol}

To verify the \gls{rf} geometry in \fref{fig:junction_spline} and \tref{tab:junction_final},
we performed an \gls{fem} simulation\footnote{
Using Comsol Multiphysics 5.5
} of its \gls{rf} field, taking into account of the extra \um{5} gaps between electrodes imposed by the fabrication process.
In the \gls{fem} simulation, we kept the \gls{rf} electrodes the same as the \textit{SurfacePattern} model described in last section,
and subtracted the width of the gaps from the \gls{rf}-grounded electrodes.
For instance, in the linear regions, the width of the middle control electrode $w_\textrm{g}$ is reduced from \um{41.5} (\tref{tab:junction_linear}) to \um{31.5}
(in future trap designs, a better approach would be to subtract half the gap width from both neighboring electrodes, as discussed in the appendix of \cite{House2008}.)
In what follows, the pseudo-potentials are calculated for $^{40}\textrm{Ca}^+$ with \SI{40}{\volt}\textsubscript{peak} \gls{rf} voltage at $\Omega_\textrm{RF}=2\pi\times\MHz{40}$ applied to the \gls{rf} electrodes.

\subsubsection{Transport path with minimal pseudo-potential}
\begin{figure}[htb]
    \centering
    \includegraphics[width=3.0in]{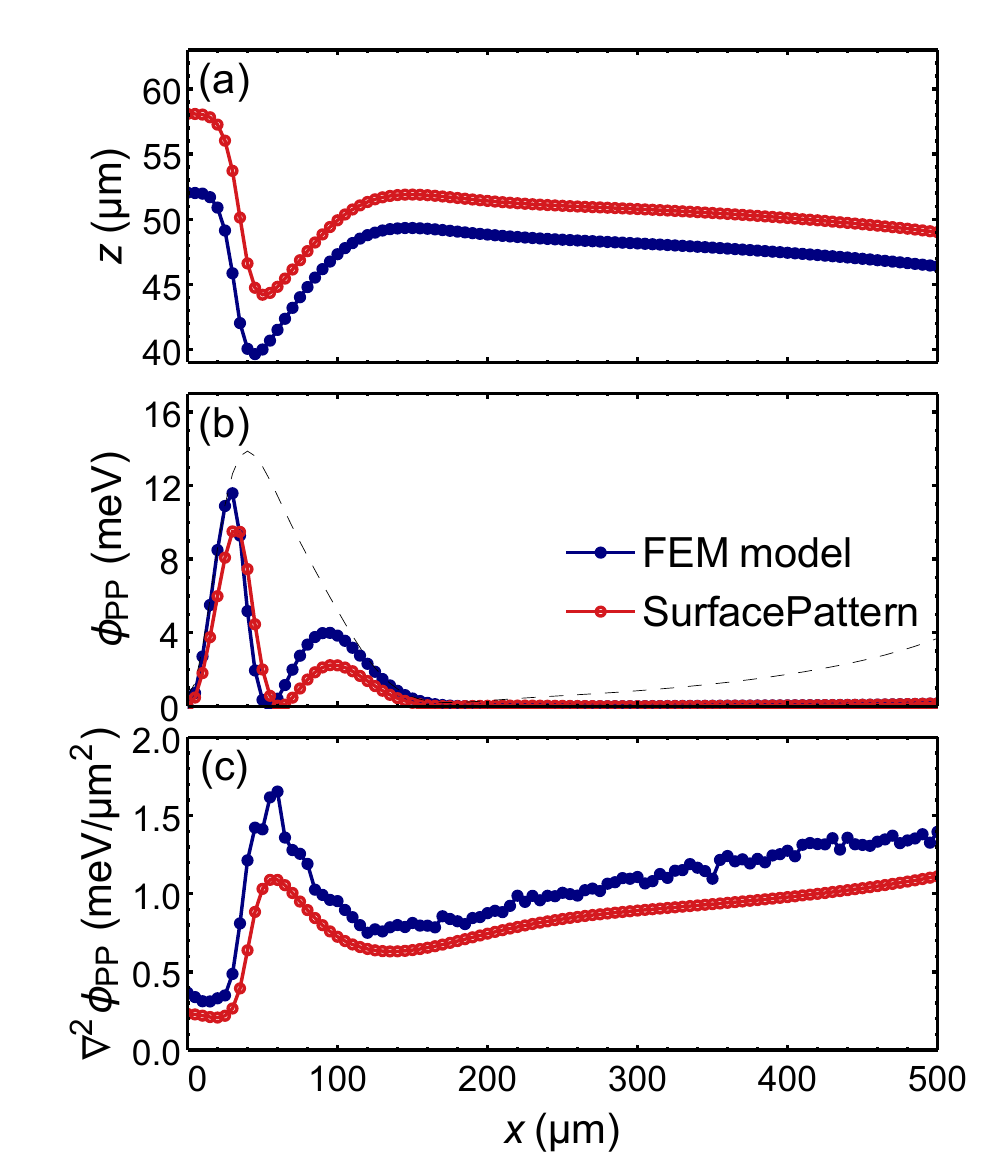}
    \caption{
    The transport path with minimal pseudo-potential (a), the corresponding pseudo-potential (b), and the pseudo-potential confinement (c) along this path. 
    The red traces are for the \textit{SurfacePattern} model, on which the numerical optimization is based.
    The blue traces are evaluated with the \gls{fem} simulation.
    The dashed line in (b) shows the pseudo-potential at constant height $z=\um{50}$ (FEM model) for comparison, the same as \fref{fig:junction_fixedH}(a).}
    \label{fig:path_minPP}
\end{figure}

\Fref{fig:path_minPP}(a) shows the transport path in which the ion is always placed at the corresponding pseudo-potential minimum for each axial position.
\Fref{fig:path_minPP}(b) and \fref{fig:path_minPP}(c) show the corresponding pseudo-potential and the pseudo-potential confinement along this transport path.
In the linear region, 
the \gls{fem} model predicts lower heights of the pseudo-potential minima by about $\um{2.7}$ and stronger pseudo-potential confinements than the \textit{SurfacePattern} model.
These discrepancies are due to the \um{5} gaps between the electrodes, which are only present in the \gls{fem} simulation.
We confirmed this with an \gls{fem} simulation of a simpler linear trap with the same \gls{rf} geometry as in \tref{tab:junction_linear}.
When the gaps between electrodes are reduced to \um{1}, 
the \gls{fem} model agrees with the \textit{SurfacePattern} model to within 0.7\% in the height of pseudo-potential minima,
and within 0.1\% in the pseudo-potential confinement evaluated at \um{50} above the trap surface.
Close to the junction center, the discrepancy between the two models in \fref{fig:path_minPP}(a) becomes larger.
This is likely due to the stronger contribution from the gaps, 
as the spacing between the \gls{rf} electrodes is narrower than in the linear region whereas the gap is fixed at \um{5}.
The small slope $\dd z/\dd x$ in the linear region of \fref{fig:path_minPP}(a) is due to both the junction structure around $x=0$ and the termination of the \gls{rf} electrodes at $x=\um{750}$.

Approaching the junction center ($\um{35}\leq x\leq\um{100}$), lowering the ion height can reduce the pseudo-potential, 
as shown in \fref{fig:path_minPP}(b),
where the pseudo-potential at the constant height of \um{50} 
is drawn in dashed line for easier comparison. 
Close to the junction center, the ion has to be lifted above the target trap height $h$ in order to achieve the minimal pseudo-potential.

The minimal pseudo-potential confinement along this transport path is 
\SI{0.2}{\milli\eV\per\micro\meter^2}, as shown in \fref{fig:path_minPP}(c). 
If the anti-confinement of a $\MHz{1.5}$ axial static trapping potential is equally distributed to the two radial modes,
this confinement allows for $\MHz{2.29}$ radial modes.
Therefore, this \gls{se} junction trap is fully confining and provides sufficient confinement to maintain a $\MHz{1.5}$ axial frequency, 
even when one opts for the transport path with minimal pseudo-potential.
Other transport strategies that are more favorable for the overall confinement are presented in the next sections.

\begin{figure}[htb]
    \centering
    \includegraphics[width=360pt]{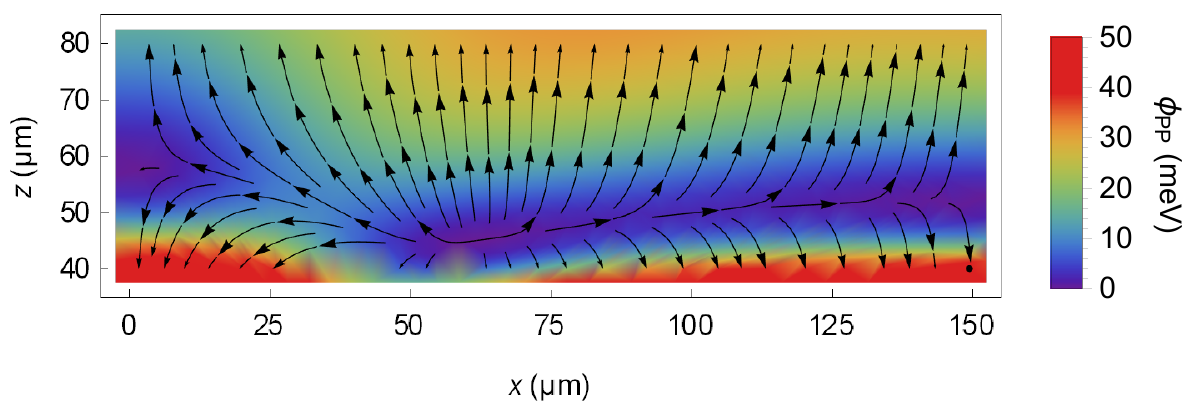}
    \caption{The direction of \gls{rf} field (black streamlines) evaluated in the symmetric plane $y=0$ of the junction trap.
    Because of the mirror symmetry, the $y$-component of the \gls{rf} is zero in this plane,
    so only the $x$ and $z$ components are plotted.
    The background color indicates the pseudo-potential.}
    \label{fig:junction_rf_field}
\end{figure}

For a better understanding of the pseudo-potential profile in \fref{fig:path_minPP}(b), 
we plot the direction of the \gls{rf} field in the symmetric plane $y=0$ in \fref{fig:junction_rf_field}.
The background color indicates the pseudo-potential. 
Close to the junction center ($0<x\leq\um{63}$), the \gls{rf} field has an $x$-component towards the island of \gls{rf} ground at the junction center (see \fref{fig:junction_spline}).
On the other hand,
far enough from the junction center, the \gls{rf} field has to point outwards. 
This is because the electric potential created by a certain electrode at the ion position is proportional to the solid angle seen from the ion~\cite{Wesenberg2008,BiotSavartES2001},
and the area of the \gls{rf} electrode is larger at the junction center than at the end of the linear arm.
In \fref{fig:junction_rf_field}, the cross-over between the inward and outward \gls{rf} fields happens at $x=\um{63}$.
Hence there is a pseudo-potential null at $x=\um{63}$ in \fref{fig:path_minPP}(b) splitting the pseudo-potential barrier into two peaks.
The \gls{rf} field has the largest horizontal component in the region $\um{25}<x<\um{40}$, 
corresponding to the pseudo-potential barrier shown in \fref{fig:path_minPP}(b).
Closer to the junction center, the horizontal \gls{rf} component reduces due to the cancellation from the opposite arm at $x<0$.
At the junction center, the horizontal \gls{rf} field is zero due to symmetry, 
so there exists a height with zero pseudo-potential.

To estimate the influence of the pseudo-potential barrier on ion transport,
we evaluated the ion motion at $(x,y,z)=(\um{30},0,\um{53.7})$, 
corresponding to the peak of pseudo-potential barrier shown in \fref{fig:path_minPP}(b).
After finding the principal axes of the \gls{rf} potential by diagonalizing its Hessian matrix,
we found that the \gls{rf} field is primarily along the direction of the least pseudo-potential curvature (\SI{0.26}{\milli\volt\per\micro\meter^2}), 
with a projection of \SI{3.2e4}{\volt/\meter}. The projections of the \gls{rf} field along the orthogonal directions are smaller by more than one order of magnitude.
We therefore focused on the 1-dimensional \gls{emm}~\cite{berkeland1998micromotion} driven by the non-vanishing \gls{rf} field along this direction.

Keeping the electric potential up to quadrupole terms, the ion motion driven by the non-vanishing \gls{rf} field is described by the inhomogeneous Mathieu equation~\cite{berkeland1998micromotion}.
If a static potential is applied to create $\omega_0=2\pi\times\MHz{1.5}$ trap frequency along the direction under investigation,
the pseudo-potential curvature along this direction is negligible compared to the curvature of the static one, 
similar to the axial mode in a standard linear trap.
We can thus simplify the 1-dimensional ion motion in this direction as a driven harmonic oscillator 
\begin{eqnarray}
    m\dv[2]{r}{t}+ m\omega_0^2 r
    =Q E_\textrm{RF}\cos(\Omega_\textrm{RF}t),
    \label{eqn:approx_rf_driven_motion}
\end{eqnarray}
whose particular integral can be obtained using the technique of variation of parameters~\cite{McLachlan51,richards1983analysis} as
\begin{eqnarray}
    \fl
    r(t) = \frac{Q E_\textrm{RF}}{m(\omega_0^2-\Omega_\textrm{RF}^2)}\cos(\Omega_\textrm{RF}t)
    \approx-\frac{Q E_\textrm{RF}}{m\Omega_\textrm{RF}^2}\cos(\Omega_\textrm{RF}t)
    = -\sqrt{\frac{4Q\phi_\textrm{PP}}{m\Omega_\textrm{RF}^2}}\cos(\Omega_\textrm{RF}t),
    \label{eqn:junction_emm}
\end{eqnarray}
which predicts \um{1.2} amplitude of the \gls{emm}.
For comparison, the \gls{emm} amplitude at the peak of pseudo-potential barrier in \cite{Blakestad2011} is estimated to be \um{6.9}\footnote{
The relevant parameters are: $^9$Be$^+$ ion, \SI{0.3}{\eV} pseudo-potential, and $\MHz{83}$ \gls{rf} drive.
}.

\subsubsection{Transport with fixed height}
\begin{figure}[htb]
    \centering
    \includegraphics[width=3.0in]{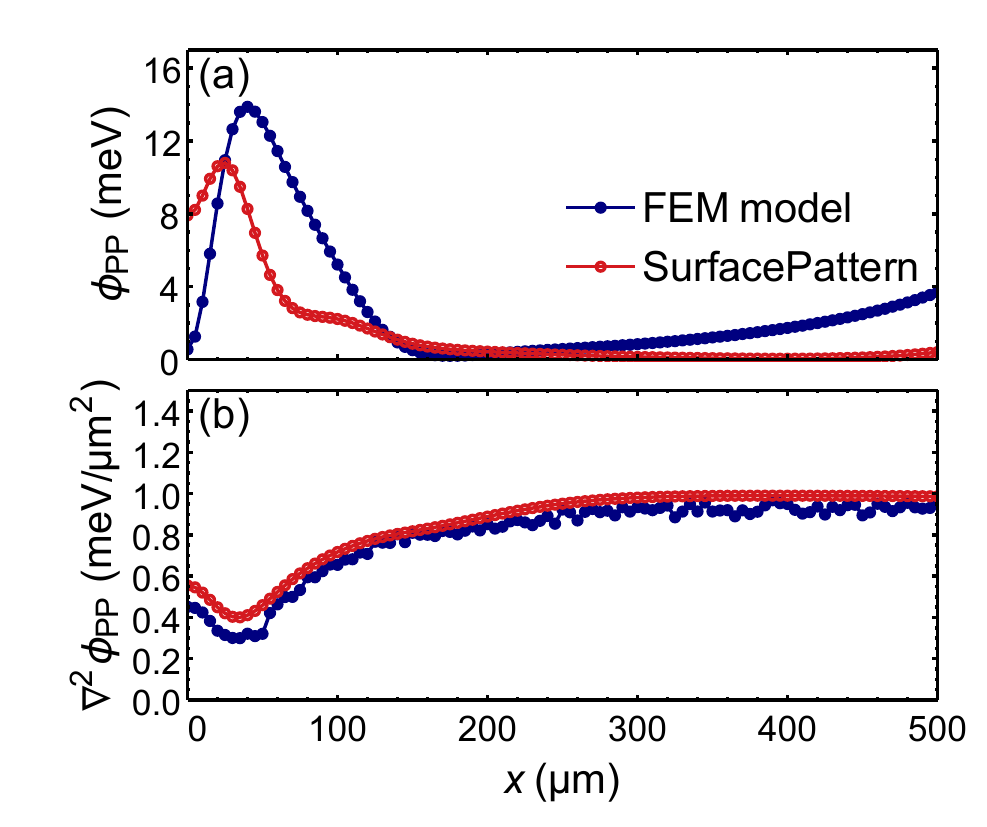}
    \caption{The pseudo-potential (a) and the total pseudo-potential confinement (b) evaluated at a fixed height of \um{50} as a function of axial displacement from the junction center $x=0$.
    }
    \label{fig:junction_fixedH}
\end{figure}

If the ion is transported at a fixed height of \um{50} above the trap surface,
the pseudo-potential and pseudo-potential confinement are shown in \fref{fig:junction_fixedH} as functions of the axial displacements from the junction center $x=0$.
The ion experiences larger total pseudo-potential confinement than in the path in \fref{fig:path_minPP}(a), at the price of higher pseudo-potential barrier.
The minimal pseudo-potential confinement along this ion path (\SI{0.3}{\milli\eV\per\micro\meter^2}) allows for a \MHz{1.5} axial mode and two \MHz{2.84} radial modes, 
assuming the confinement is equally distributed between the two radial modes.

The discrepancy in pseudo-potentials in \fref{fig:junction_fixedH}(a) between the two models is mainly a consequence of the different positions of the pseudo-potential minima shown in \fref{fig:path_minPP}(a).
The evaluation height, $z=\um{50}$, is closer to the pseudo-potential minima of the \textit{SurfacePattern} model, 
in particular towards the end of the junction arms.
In \fref{fig:junction_fixedH}(b), the pseudo-potential confinements predicted by the two models show a good agreement when evaluated at the same height. 

For comparison, in an optimization where the cost term $f_1$ for pseudo-potential confinement is made to be dominant by setting $w_1=10 w_2$ in \eref{eqn:dual_cost}, the minimal pseudo-potential confinement along the constant-height transport path for the optimization output (\SI{0.67}{\milli e\volt/\micro\meter^2}) is more than two times higher than in \fref{fig:junction_fixedH}(b), but the maximal $\phi_\textrm{PP}$ increases to \SI{58}{\milli e\volt}.
On the other hand, the geometry optimized mainly for the minimal pseudo-potential gradient (for $w_2=10w_1$) leads to a lower pseudo-potential barrier (\SI{3.8}{\milli e\volt}) than in \fref{fig:junction_fixedH}(a) whereas the minimal pseudo-potential confinement decreases to \SI{0.1}{\milli e\volt/\micro\meter^2}.

\subsubsection{Transport path with constant confinement}
\begin{figure}[htb]
    \centering
    \includegraphics[width=3in]{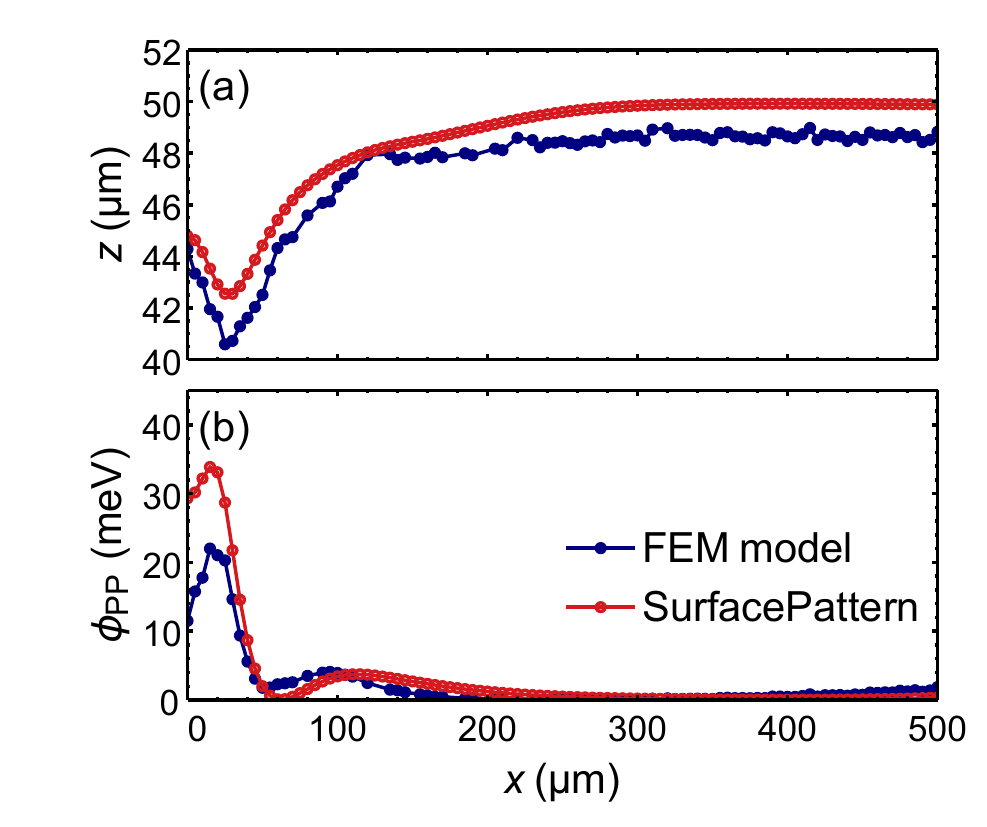}
    \caption{
    (a) A transport path maintaining a constant pseudo-potential confinement of \SI{1.0}{\milli\eV\per\micro\meter^2}. 
    (b) The pseudo-potential along this transport path.
    }
    \label{fig:path_constConf}
\end{figure}

Another strategy of ion transport is to maintain a constant pseudo-potential confinement along the path.
The transport paths shown in \fref{fig:path_constConf}(a) maintain a pseudo-potential confinement of \SI{1.0}{\milli\eV\per\micro\meter^2}. 
Approaching the junction center,
the ion height has reduced by $\sim\um{7}$ to maintain the same confinement, 
which is qualitatively the same observation as in \cite{Wright2013GTRI_junction}.
\Fref{fig:path_constConf}(b) shows the pseudo-potential along this path.
The ion will experience a higher pseudo-potential (\SI{33.9}{\milli\eV}),
which corresponds to \um{2.28} \gls{emm} amplitude estimated using \eref{eqn:junction_emm}.

Because of the larger deviation from the pseudo-potential minima in \fref{fig:path_constConf}(a) than in the previous two transport paths,
the validity of the pseudo-potential approximation becomes one concern.
Within the pseudo-potential approximation, the electric potentials are expanded up to 2nd order,
but when the ion is far enough from the saddle point of the electric potentials,
the higher order anharmonic terms could have non-negligible influence on the ion motion.
To verify the ion confinement along the transport path in \fref{fig:path_constConf} and to better estimate the amplitude of the ion's driven motion,
we numerically calculated the ion trajectory at the position of largest pseudo-potential $(x,y,z)=(\um{15},0,\um{43.52})$ without making the pseudo-potential approximation. 

First of all, we diagonalized the Hessian of the pseudo-potential at this position to find its principal axes, 
which are listed in \tref{tab:eigensystem_maxPP} together with the corresponding eigenvalues and the projections of the \gls{rf} field along these directions.
We then optimized the static voltages applied to the control electrodes (whose layout will be introduced in \sref{sec:junction_control})
to redistribute the confinement,
which results in $(6.51,4.03,1.5)~\si{\mega\hertz}$ secular frequencies along the three principal axes in \tref{tab:eigensystem_maxPP}.
To keep the time-averaged ion position at the specified position, the static voltages were used to compensate the ponderomotive force due to the gradient of the pseudo-potential.
The largest magnitude of the static voltage involved is \SI{2.48}{\volt}. 

\begin{table}[htp]
\centering
\caption{Principal axes of the pseudo-potential Hessian, the corresponding eigenvalues, the secular frequencies, and the \gls{rf} electric field at $\vec{r}=(\um{15},0,\um{43.52})$.}  
\label{tab:eigensystem_maxPP}
\begin{indented}
\lineup
\item[]
\begin{tabular}{@{}llll@{}} 
\br
{Principal axis $(x,y,z)$}& \makecell{Eigenvalue\\  (\si{\milli\eV\per\micro\meter^2})}& \makecell{Secular frequency\\(MHz)} & \makecell{Electric field\\(V/m)}\\ 
\mr
$(0.195,0,0.981)$& 0.991 &$7.79$ & 53218\\
$(0,1,0)$& 0.104& $2.53$ & \0\0188\\
$(-0.981,0,0.195)$&\-0.099 & --- & \-28627 \\
\br
\end{tabular}
\end{indented}
\end{table}

\begin{figure}[htb]
    \centering
    \includegraphics[width=3in]{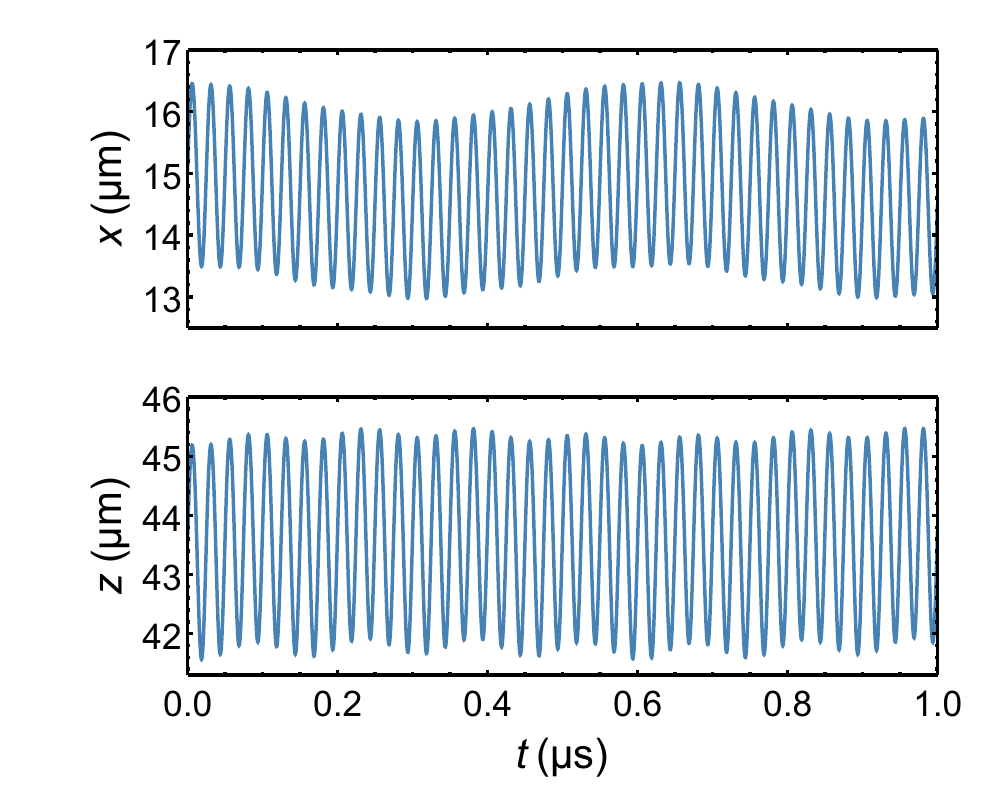}
    \caption{Numerical simulation of the ion trajectory in the presence of a strong \gls{rf} field close to junction center.
    The static voltages are designed to keep the time-averaged ion position at $\vec{r}=(\um{15},0,\um{43.52})$, where the pseudo-potential is largest in \fref{fig:path_constConf}.}
    \label{fig:trajectory_numerical}
\end{figure}

With this set of static voltages, we numerically solved the equation of ion motion
\begin{eqnarray}
    m\Ddot{\vec{r}} = Q\qty[\vec{E}_\textrm{st}(\vec{r}) + \vec{E}_\textrm{RF}(\vec{r})\cos(\Omega_\textrm{RF} t)]
\end{eqnarray}
using the analytical expressions for $\vec{E}_\textrm{st}$ and $\vec{E}_\textrm{RF}$ from \textit{SurfacePattern}.
The $x$ and $z$ components of the solution are shown in \fref{fig:trajectory_numerical}.
Due to the strong \gls{rf} field, the $\MHz{40}$ \gls{emm} is the dominant motional component.
The modulation of the \gls{emm} envelop is due to the secular motion.
The \gls{emm} amplitude obtained from the numerical integration is \um{2.34}, 3\% larger than the estimation from \eref{eqn:junction_emm}.
Hence, the pseudo-potential approximation remains a good approximation in this extreme case.

\subsection{Comparison with a simple junction}
\begin{figure}[htbp]
    \centering
    \includegraphics[width=4in]{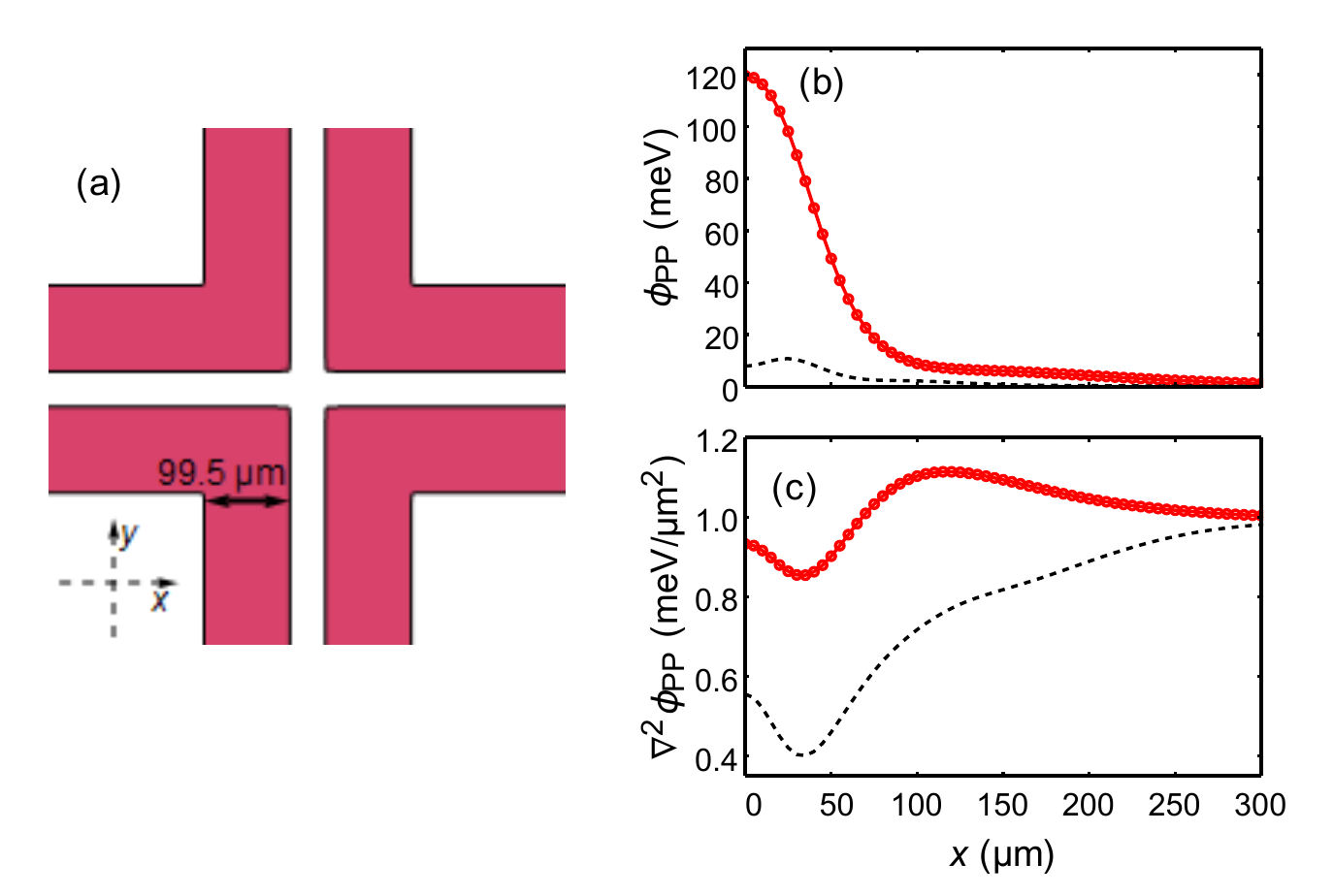}
    \caption{
    (a) An X-junction constructed by simply crossing two linear traps, each with \um{99.5} wide \gls{rf} electrodes spaced by \um{41.5}, same as in \tref{tab:junction_linear}.
    The pseudo-potential (b) and the total pseudo-potential confinement (c) are evaluated using \textit{SurfacePattern} at a fixed height of \um{50}. 
    The same curves as in \fref{fig:junction_fixedH} for the optimized junction geometry are shown in dashed lines for comparison.
    }
    \label{fig:naive_junction}
\end{figure}

To benchmark the optimized junction geometry, we can compare it with a simple junction constructed by crossing two linear traps in right angle, as shown in \fref{fig:naive_junction}(a).
The \gls{rf} geometry of the linear traps are the same as listed in \tref{tab:junction_linear}.
The pseudo-potential and the total pseudo-potential confinement evaluated at fixed height of \um{50} are plotted in \fref{fig:naive_junction}(b-c).
The maximal pseudo-potential in \fref{fig:naive_junction}(b) is one order of magnitude higher than the optimized geometry as in \fref{fig:junction_fixedH}.

\begin{figure}[htb]
    \centering
    \includegraphics[width=3in]{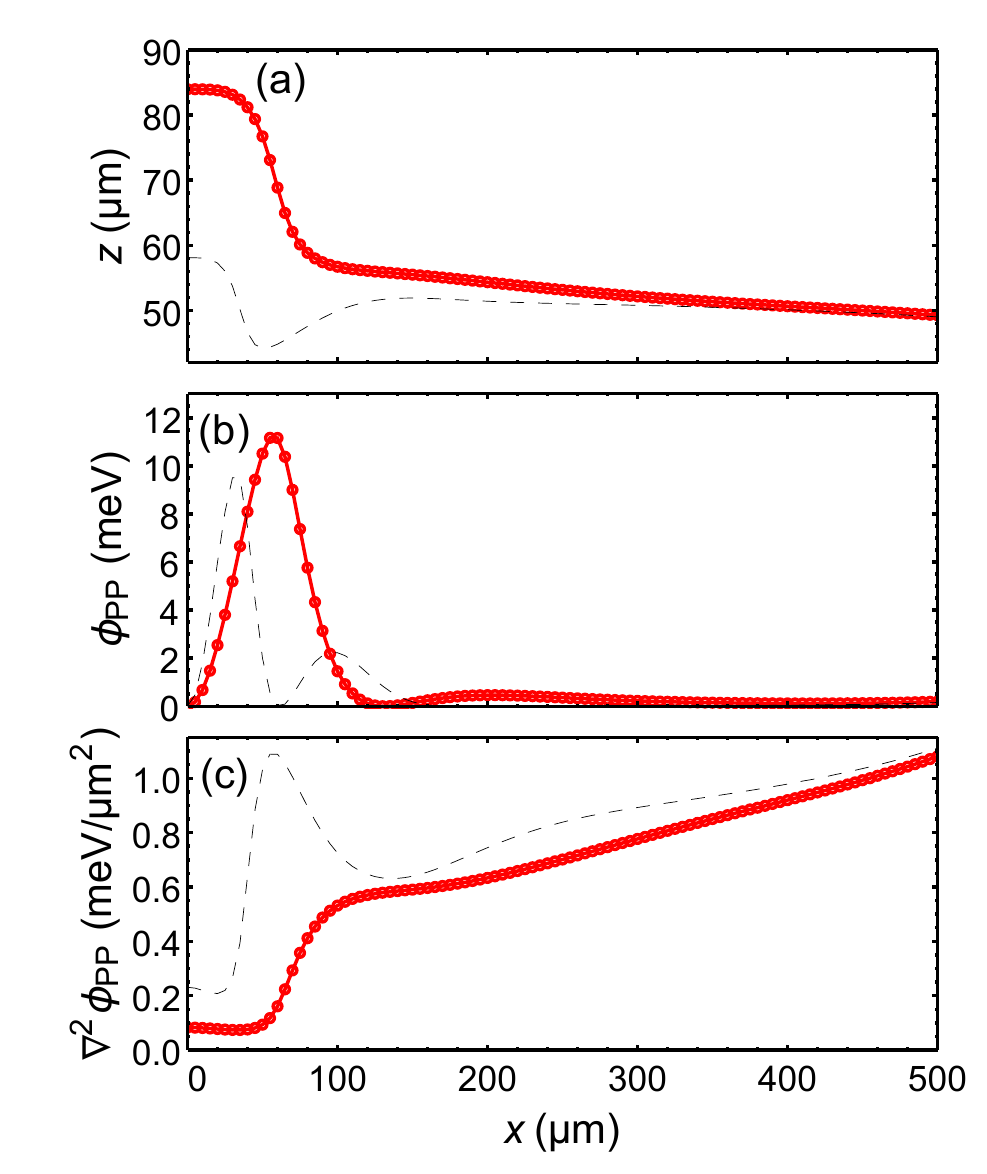}
    \caption{(a) A transport path with minimal pseudo-potential for the simple junction trap shown in \fref{fig:naive_junction}(a),
    the corresponding pseudo-potential (b), and the pseudo-potential confinement (c) along this path, simulated with \textit{SurfacePattern}.
    For comparison, the dashed lines show the same curves as in \fref{fig:path_minPP} for the optimized junction geometry.
    }
    \label{fig:naive_minPP}
\end{figure}

To keep a minimal pseudo-potential along the transport path,
the height of the ion has to be lifted up to \um{84} (\fref{fig:naive_minPP}(a)),
which reduces the maximal pseudo-potential along the transport path by one order of magnitude (\fref{fig:naive_minPP}(b)),
to roughly the same level as the optimized geometry in \fref{fig:path_minPP}(b).
However, as the ion is lifted higher above the trap surface, 
the minimal available pseudo-potential confinement along this path drops to \SI{0.07}{\milli\eV\per\micro\meter^2},
which is about 30\% that of the optimized geometry in \fref{fig:path_minPP}(c).
If one motional mode is kept at $\MHz{1.5}$, such a total confinement only allows for $\MHz{1.06}$ frequency on the other two modes.

\subsection{Final layout of RF electrode}
\begin{figure}[htb]
    \centering
    \includegraphics[width=3.5in]{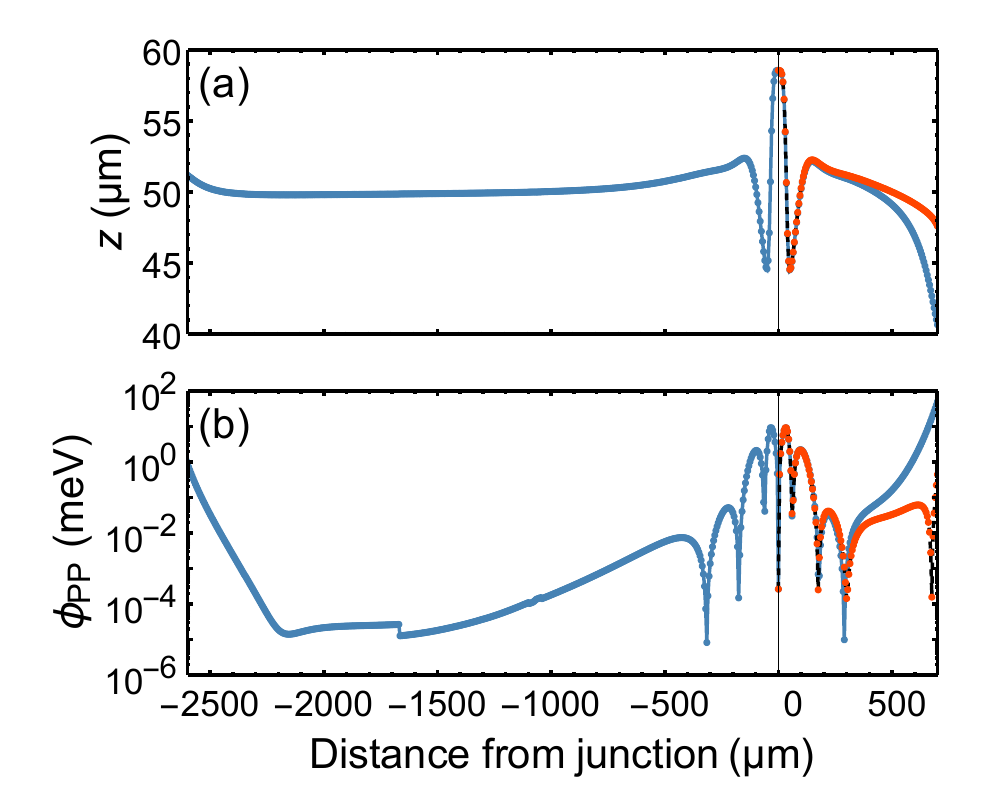}
    \caption{Height of the pseudo-potential minima (a) and the corresponding pseudo-potential (b) for the final \gls{rf} electrode geometry, evaluated in the symmetry plane of each arm with \textit{SurfacePattern}.
    The negative displacements are along the long arm.
    The wirebonding pad at \um{-2700} is responsible for the rise of pseudo-potential on that end.
    There is an \nm{8} discontinuity at \um{-1665} in (a) due to numerical imprecision, which causes the discontinuity in (b). 
    The positive displacements are along the \um{750} short arms. 
    The orange points are for the two arms with the arc interconnection structure,
    which flattens the pseudo-potential profile compared to the straight termination.
    }
    \label{fig:junction_field_final}
\end{figure}

In the final layout of the \gls{rf} electrode (see \fref{fig:junction_DC} for the overall layout also including the control electrodes), we chose three arms of the junction trap to be \um{750} long, 
and elongated one arm to \mm{2.7}.
The long arm allows for a working zone far from the junction to reduce the residual axial \gls{rf} field from the pseudo-potential barrier.
The wirebonding pad for the \gls{rf} electrodes is placed at the end of the long arm,
and thus it is also preferable to make this arm long to minimize the extra \gls{rf} field from the wirebonds in the working regions. 
The height of pseudo-potential minima and the corresponding pseudo-potential are plotted in \fref{fig:junction_field_final}.
The pseudo-potential of \SI{1.8e-4}{\milli\eV} at \um{-1000} would give rise to \nm{5.2} \gls{emm}.

Although it is technically feasible to make separate electrical connections to the four \gls{rf} electrodes using vias and other metal layers underneath the trap surface,
we decided to interconnect them using the top metal layer on this trap chip.
This ensures low impedance of the connections and minimizes the phase difference of the \gls{rf} fields from different \gls{rf} electrodes.
The second available metal layer in our fabrication technology (see \sref{sec:chip_layout}) is \gls{rf} grounded and shields the silicon substrate of the trap chip from the \gls{rf} tracks, 
reducing the \gls{rf} loss~\cite{GTRI2012} and potential laser-induced photo-effects due to the photoionization of atoms in the substrate by the UV lasers involved in the experiments~\cite{Mehta2014}.

The two long \gls{rf} electrodes are connected by the same wire bonding pads.
To connect the other two \gls{rf} electrodes with minimal variation in \gls{rf} impedance,
we made arc structures with the same width as the \gls{rf} electrodes (see \fref{fig:junction_DC}),
which keeps the width of this interconnection piece constant.
As can be seen from \fref{fig:junction_field_final}, the arc interconnection structures also flatten the pseudo-potential profile close to the termination of the \gls{rf} electrodes (orange points).

\section{Control electrodes}
\label{sec:junction_control}
After finalizing the geometry of the \gls{rf} electrodes, we patterned the control electrodes for creating the quasi-static trapping potential and for transporting the ion according to the shape of the \gls{rf} electrodes.

To keep a fixed \um{5} gap from the non-trivial outline of the \gls{rf} electrodes,
we first sampled the spline functions defining the \gls{rf} electrodes at 100 points, 
with about \um{1} spacing between each two.
For each line segment connecting the neighbouring sample points, 
we calculated its transverse direction in the trap plane, along which we shifted the line segment by \um{5}.
Finally, we interpolated the shifted sample points to obtain the outline of the control electrodes. 

\subsection{Segmented middle electrodes}
\label{sec:seg_middle}
Leveraging the flexibility in electrical connections using the second metal layer (see \sref{sec:chip_layout}), 
we segmented the middle electrodes to realize the 5-wire geometry of \gls{se} traps~\cite{Wesenberg2008,Chiaverini2005},
which has two advantages over the segmentation of the outer two wires.
The number of control electrodes is roughly halved for achieving the same level of flexibility in ion transport along the trap axes.
Furthermore, since the middle segments are closer to the ions, 
they can create stronger curvatures at the ion position with a fixed voltage~\cite{Wesenberg2008}. 
To create $\MHz{1}$ axial frequency, the outer segmentation method would require $\sim 20$ times higher voltages than the middle segmentation for our \gls{rf} geometry.

\begin{figure}[htb]
    \centering
    \includegraphics[width=4.5in]{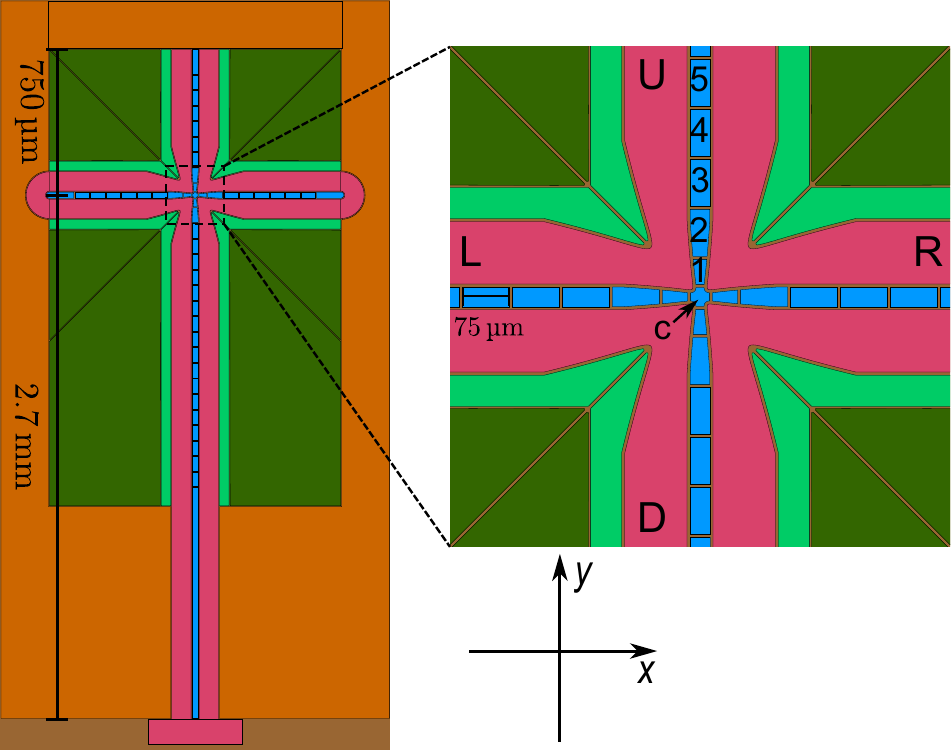}
    \caption{\gls{rf} (red) and control electrodes around the junction center.
    The middle control electrodes (blue) are divided into \um{75} segments, except for the \um{30} one at junction center (labeld ``c'') and the four \um{40} electrodes around it (indexed ``1'').
    The junction arms are labeled ``U'', ``D'',``L'' and ``R''.
    Each middle electrodes is indexed by the arm it belongs to and its distance from junction center, e.g. U3.
    The outer control electrodes are segmented into 4 pieces in each quadrant (light and dark green).
    The rest of the trap surface is covered by a ground plane, colored in dark yellow.
    The uncovered space at the bottom are for wirebonding pads for the control electrodes.
    The gap between electrodes is \um{5}.}
    \label{fig:junction_DC}
\end{figure}

The segmented middle electrodes are shown in blue in \fref{fig:junction_DC}.
When operating the trap, smaller segments are preferable as they allow for fine adjustments of the trapping potential.
However, the number of control electrodes is limited by the space for their wirebonding pads on the perimeter of the trap chip,
which imposes a lower bound to the minimal length of each control electrodes. 
Due to this limitation, we divided the middle control electrodes into segments of \um{75} length in the linear regions.
For more flexibility in the junction transport, we reduced the length of the 4 control electrodes adjacent to the center to \um{40}.
The control electrode at the center is a \um{30} wide square, whose corners are removed due to the extrusion of the \gls{rf} electrodes.
The width of the middle control electrodes in the linear region is \um{31.5}, 
whereas in vicinity of the junction center they have variable widths following the outline of the \gls{rf} electrodes.
There are overall 48 segmented middle electrodes, with 9 on each short arm, 20 on the long arm, and 1 at the junction center.

As a verification of the segmented middle electrodes, we calculated the voltages required to create $\MHz{1.5}$ axial frequency along the minimal pseudo-potential transport path in \fref{fig:junction_field_final}(a).
All the required voltages are within \SI{-2}{\volt} to \SI{5}{\volt} and should be straightforwardly obtainable from typical \glspl{dac}.

\subsection{Outer control electrodes}
Doppler cooling of the ion motion requires the wave vector of the laser beam to have non-zero projections along the modes to be cooled, 
and that the trap frequencies are non-degenerate~\cite{amini2011microfabricatedChipTrap}.
In the linear region of the trap, the two radial modes are parallel and perpendicular to the trap surface respectively when the trapping potential possesses mirror symmetry about the trap axis.
Hence, for a Doppler cooling laser beam running parallel to the trap surface,
the principal axes of the trapping potential have to be rotated to cool the out-of-plane mode. 
To rotate the principal axes of the trapping potential, differential voltages need to be applied onto the outer electrodes to break the mirror symmetry about the trap axis.
Furthermore, to simultaneously keep the potential minimum on the trap axis,
the outer electrodes must be segmented transversely (see \fref{fig:junction_DC}).

The junction structure automatically 
tilts the principal axes of the pseudo-potential and lifts the degeneracy of motional modes in its vicinity (see e.g. \tref{tab:eigensystem_maxPP}).
At the junction center, the central-most control electrodes can tilt the principal axes more effectively than the outer ones.
Hence, the geometry of the transverse segmentation of the outer electrodes is most important for the operation at the linear regions of the trap.
Therefore, to optimize this transverse segmentation, we simulated a simple linear trap with the \gls{rf} geometry shown in \tref{tab:junction_linear}.

From the simulation, we found that lifting the degeneracy of the radial modes is more demanding than merely rotating the principal axes.
Thus we fixed the rotation angle to \ang{20} and used the attainable maximal radial mode splitting to benchmark different segmentation ratios of the outer electrodes. 
For a fixed voltage range from \SI{-10}{\volt} to \SI{10}{\volt}, we adjusted the segmentation ratio such that the thinner (light green in \fref{fig:junction_DC}) and the wider (dark green in \fref{fig:junction_DC}) outer electrodes reach their limiting values roughly at the same time when creating the maximal mode frequency splitting.
In the final geometry shown in \fref{fig:junction_DC}, the thinner outer electrodes are \um{49.75} wide at the linear region, whereas the wider ones are \um{580} wide.
We can intuitively understand such a ratio from the fact that the thinner electrodes are closer to the ion and thus are relatively more influential, 
whereas the segments far from the trap axis need a larger area to make a comparable impact on the ion. 
Using $\SI{\pm 10}{\volt}$, such a geometry can create about \kHz{700} splitting  
for degenerate radial modes at $\MHz{6.06}$, 
resulting in two radial modes at $\MHz{6.24}$ and $\MHz{5.54}$ with the radial principal axes rotated to \ang{20} and \ang{70} to the $z$-axis.

It is worth noting that this final geometry of outer electrodes is obtained heuristically from simulation. 
The analytical expressions for the electric field from \gls{se} in \cite{House2008} can potentially be used to find the theoretically optimal geometry.

\subsection{Transport waveform}
\begin{figure}[htb]
    \centering
    \includegraphics[width=4in]{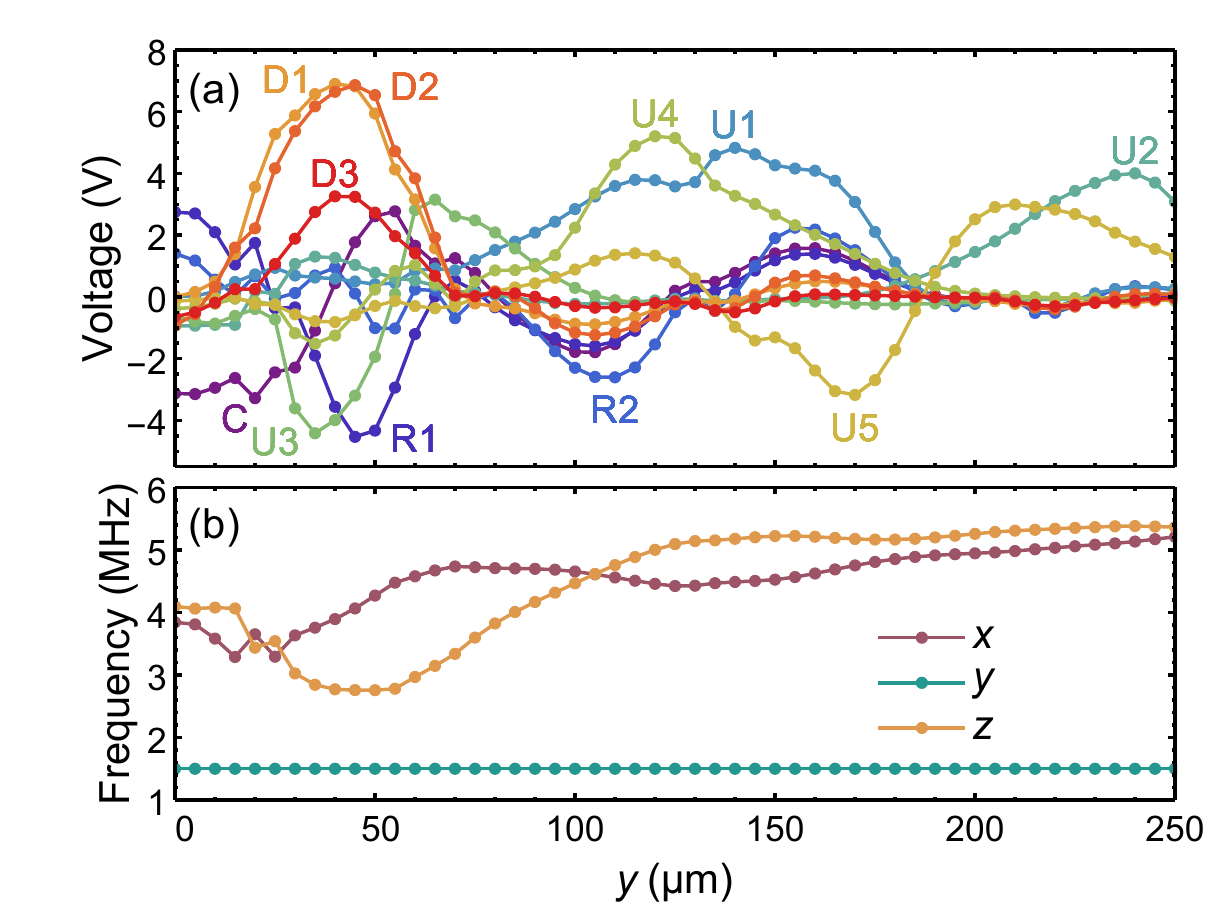}
    \caption{A voltage waveform of junction transport (a) and the resulting secular frequencies (b).
    The full waveform transports the ion from the linear region of the upper arm ($y=\um{500}$) to the long arm on the other side of the junction ($y=-\um{95}$).
    For clarity, only part of the waveform and only the voltages exceeding \SI{\pm2}{\volt} are shown. 
    The voltages applied to arm L are identical to those on arm R.
    }
    \label{fig:waveform}
\end{figure}

To verify that junction transport can be realized within our feasible voltage range, 
we optimized a transport waveform, as shown in \fref{fig:waveform}.
To obtain a smooth waveform, the voltage changes between subsequent steps are penalized, 
which was not taken into account in the voltage checks at individual axial positions performed in \sref{sec:seg_middle}.

The waveform generation is based on the Python library CVXPY with the MOSEK solver backend.
We formulated the optimization problem using the electric fields and curvatures at the ion positions.
We discretized the transport path into a set of $N$ positions $\qty{\vec{r}_i},i\in [1,N]$,
at which a sequence of static trapping wells are described. 
The trap model based on \textit{SurfacePattern} was used for the waveform generation.
For each control electrode $j$, we exported to the optimizer its electric field 
\begin{eqnarray*}
    \qty{E_{x}^{(j)}(\vec{r}_i),E_{y}^{(j)}(\vec{r}_i),E_{z}^{(j)}(\vec{r}_i)
    }
\end{eqnarray*}
and the 6 elements of its Hessian
\begin{eqnarray*}
   \qty{
    \mathcal{H}_{xx}^{(j)}(\vec{r}_i),\mathcal{H}_{xy}^{(j)}(\vec{r}_i),\mathcal{H}_{xz}^{(j)}(\vec{r}_i),
    \mathcal{H}_{yy}^{(j)}(\vec{r}_i),\mathcal{H}_{yz}^{(j)}(\vec{r}_i),\mathcal{H}_{zz}^{(j)}(\vec{r}_i)
    }
\end{eqnarray*}
evaluated at $\vec{r}_i$ for \SI{1}{\volt} applied voltage.
The pseudo-potential gradients $\vec{E}^\textrm{(PP)}$ and curvatures $\mathcal{H}^\textrm{(PP)}$ from the \gls{rf} electrodes were exported analogously.
The optimizer then seeks for the control voltages at each step $i$ that minimize the cost function under the constraints specified below. 
The same optimization procedure can be applied to the \gls{fem} trap model as well, 
after pre-processing the \gls{fem} simulation to construct a similar look-up table of fields and curvatures at $\qty{\vec{r}_i}$.

The constraints used in the optimization of the waveform shown in \fref{fig:waveform} are:
\begin{enumerate}
    \item Zero time-averaged field at each $\vec{r}_i$
    \begin{eqnarray}
       \vec{E}^\text{(PP)}(\vec{r}_i) + \sum_j \vec{E}^{(j)}(\vec{r}_i) = \vec{0}.
       \label{eqn:zero_total_field}
    \end{eqnarray}
    \item 
    One principal axis of the trapping potential is exactly along the trap axis, which is $y$ in this example
    \begin{eqnarray*}
        \mathcal{H}_{xy}^\text{(PP)}(\vec{r}_i) + \sum_j \mathcal{H}_{xy}^{(j)}(\vec{r}_i) = 0,\\
        \mathcal{H}_{yz}^\text{(PP)}(\vec{r}_i) + \sum_j \mathcal{H}_{yz}^{(j)}(\vec{r}_i) = 0.
    \end{eqnarray*}
    \item The axial Hessian element equals 
    the certain value $\mathcal{H}_\text{axial}(\vec{r}_i)$ that results in the specified axial frequency, \MHz{1.5} in this example
    \begin{eqnarray*}
        \mathcal{H}_{yy}^\text{(PP)}(\vec{r}_i) + \sum_j \mathcal{H}_{yy}^{(j)}(\vec{r}_i) = \mathcal{H}_\text{axial}(\vec{r}_i).
    \end{eqnarray*}
    \item Radial curvatures are above a certain limit,
    which is chosen to be $3\mathcal{H}_\text{axial}(\vec{r}_i)$ here
    \begin{eqnarray*}
        \mathcal{H}_{xx}^\text{(PP)}(\vec{r}_i) + \sum_j \mathcal{H}_{xx}^{(j)}(\vec{r}_i)  \geq 3\mathcal{H}_\text{axial}(\vec{r}_i),\\
        \mathcal{H}_{zz}^\text{(PP)}(\vec{r}_i) + \sum_j \mathcal{H}_{zz}^{(j)}(\vec{r}_i)  \geq 3\mathcal{H}_\text{axial}(\vec{r}_i).
    \end{eqnarray*}
\end{enumerate}

The cost function is formed as the weighted sum of the following contributions:
\begin{enumerate}
    \item The 1st and 2nd order time derivatives of the control voltages.
    This cost term ensures a smooth voltage waveform, reducing its high-frequency components that can cause motional excitation and impose higher technical requirements for the waveform generator hardware.
    In this preliminary waveform example in \fref{fig:waveform}, time is implicitly represented by the axial position of the ion.
    \item The distance from $\vec{r}_i$ to the active control electrodes. 
    This cost term penalizes the use of distant control electrodes, since in the \gls{QCCD} architecture parallel operations are necessary and thus the transport of the communication ion should minimally affect the potential well for the static memory ions.
    \item The residual electric field at $\vec{r}_i$. 
    Although already specified by the constraint condition \eref{eqn:zero_total_field}, 
    the outcome from numerical optimization can still have a residual electric field at the desired trapping position. 
    This term penalizes such residual field to ensure the ion is transported along the desired path.
\end{enumerate}

For the waveform shown in \fref{fig:waveform}, only the segmented middle electrodes were used to generate the potential wells.
Involving the outer electrodes in the transport results in smaller voltages fulfilling the same constraints.
We excluded them because for parallel operations these outer electrodes are responsible for maintaining the tilt of radial axes in the linear regions.
In future traps, these outer electrodes can be segmented along the axial direction to provide more flexible controls in parallel operations.
Radial mode tilting and splitting are not imposed in the preliminary waveform shown in \fref{fig:waveform},
but can be achieved independently using the redundant degrees of freedom from the outer electrodes.

\Fref{fig:waveform} confirms that the control electrode layout in \fref{fig:junction_DC} allows for a junction transport with a constant \MHz{1.5} axial frequency and at least \MHz{2.75} radial frequency.
The required voltage range \SI{-4.5}{\volt} to \SI{7.0}{\volt} can be conveniently obtained from typical \glspl{dac}.
We therefore took such a layout as the final design for the trap fabrication.

\section{Optics integrated into the junction trap} 
\label{sec:integrated_optics}
Conventionally, the laser beams for controlling the trapped ions are delivered in free space.
To avoid hitting the trap chip, for \gls{se} traps the free-space laser beams have to propagate parallel to the trap surface.
Such a geometrical constraint imposes challenges in parallel operations on a \gls{se} junction trap,
as the laser beams addressing one arm have to avoid the ions on the other arms.
For a single X-junction, it is still in principle possible to arrange the laser beams for parallel operation.
For instance, every laser beam can be aligned perpendicular to the arm it is addressing.
However, in a larger-scale trap array connected by multiple junctions,
such a strategy becomes invalid since there will be junction arms parallel to each other.
The density of usable trap modules will eventually be limited by the available optical access. 

Nanophotonics devices integrated on the trap chip offer a promising approach for scalable laser beam delivery~\cite{Mehta2016,Mehta2020,Niffenegger2020,Ivory2021}.
The laser light can be routed through waveguides inside the trap chip,
and focused to the ions by grating couplers~\cite{Mehta2017}. 
Since the laser beams are emitted from the ion trap, 
they can be directed to individual trap zones 
without affecting the parallel operation in the other zones.
Aside from the parallelizability, the integrated optics also improves the stability of light delivery since the optics are rigidly fixed to the ion trap,
and the efficiency of laser power usage due to the tighter beam foci~\cite{Mehta2020}.

On our junction traps, we used \SiN{} integrated optics to deliver the \nm{729} coherent manipulation light and the \nm{854} and \nm{866} repumping light for the $\textrm{Ca}^+$ qubits.
These junction traps are fabricated on the same wafer as our experimentally tested linear traps~\cite{Mehta2020}. 
A thin \SiN{} layer (\nm{25} thick) \um{3.5} below the ground plane extends to the edge of the trap chip.
The low-confinement optical mode it supports matches the mode of a standard single-mode fiber for optical input.
A vertical adiabatic taper couples light from the \nm{25} \SiN{} layer to a \nm{170} thick \SiN{} core separated by \nm{50} from the thinner one,
which creates a higher confinement of the optical mode for waveguide routing~\cite{Mehta2020,Mehta2019}.
After the light is routed to the trap zone, a lateral taper expands the mode size,
and a diffractive grating patterned by electron beam lithography emits the light to the ions above the trap chip~\cite{Mehta2020}.
The thickness of silicon dioxide below the gratings is chosen such that the reflection from the silicon substrate constructively interferes with the upwards-radiated light,
to enhance the out-coupling efficiency of the gratings~\cite{Mehta2017}.

\begin{figure}[htbp]
    \centering
    \includegraphics[width=3.3in]{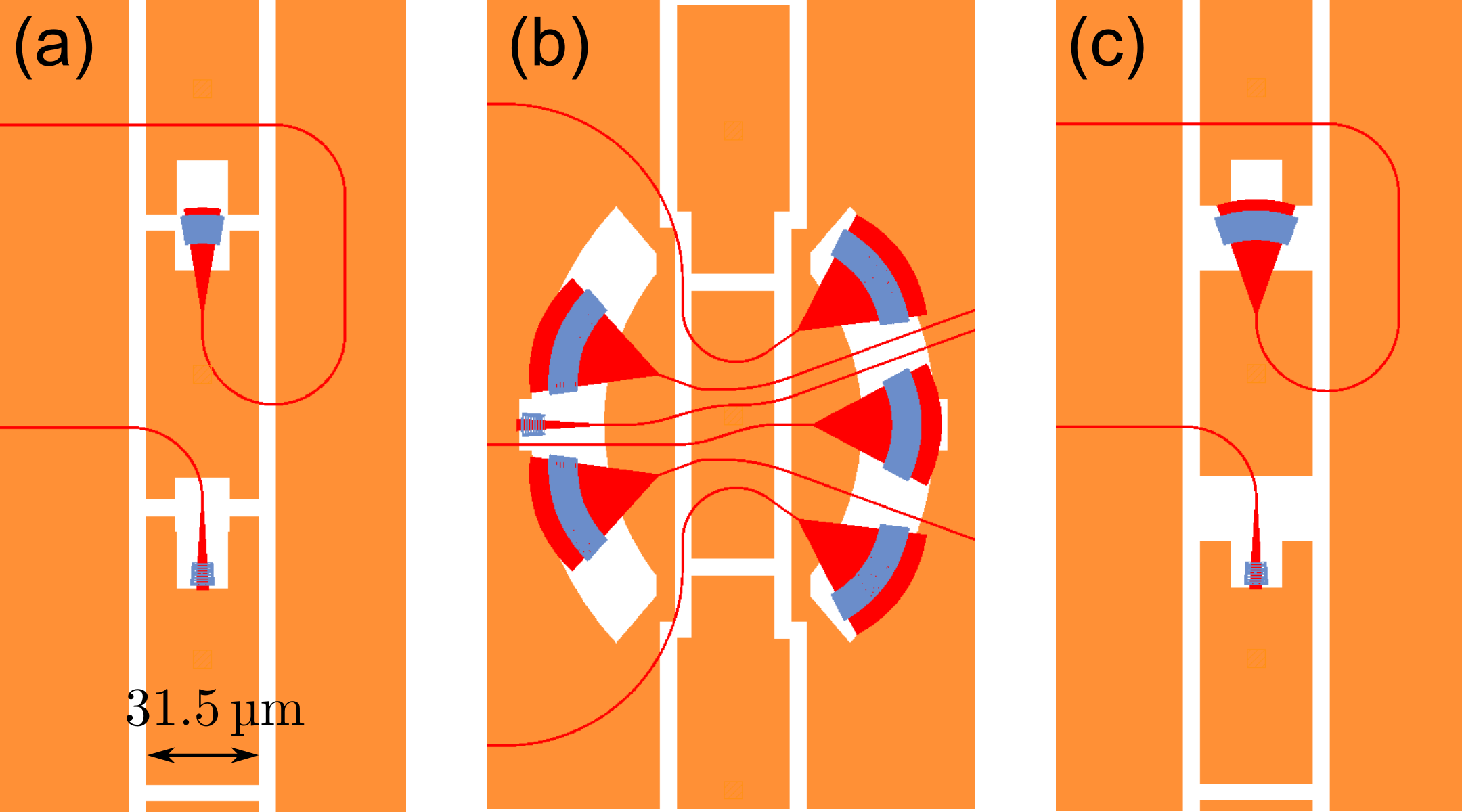}
    \caption{The grating couplers implemented on the junction traps.
    The \SiN{} waveguides and the lateral tapers for expanding the optical modes are shown in red. 
    The electron beam lithography region used for the gratings are indicated in blue. 
    The openings on the metal layers for light transmission are symmetric in each interaction zone.
    (a) shows a 2-qubit interaction zone with the upper coupler for \nm{729} and the lower for \nm{854} and \nm{866} light, 
    implemented on 4 arms of one junction trap.
    The other junction trap features a 5-ion interaction zone (b) on its long arm and the same \nm{729} individual ion addressing optics (c) on its three short arms.
    }
    \label{fig:junction_coupler}
\end{figure}

By equipping trap sites with the corresponding integrated optics, elementary units of different functionalities in quantum information processing can be realized.
Further interconnecting them with junctions enables a modular approach to large-scale trapped-ion computing.
We have made two versions of junction traps featuring integrated optics for different tasks.
On each arm of one junction trap, we placed the same two-ion interaction optics as experimentally tested in \cite{Mehta2020}, 
as shown in \fref{fig:junction_coupler}(a).
The gratings are designed to focus light to \um{50} above the trap surface.
The upper grating coupler is for \nm{729} light,
which focuses the beam to \um{3.7} Gaussian beam waist ($1/\rme^2$ radius) perpendicular to the trap axis. 
Along the trap axis, the beam waist is relaxed to \um{6.5} for illuminating two ions and driving two-qubit gates using the axial motional mode.
The lower coupler delivers the \nm{854} and \nm{866} light for laser cooling and resetting the qubit states.
The designed beam waist is about \um{20}, and hence this grating coupler takes a smaller area than the \nm{729} one.
The openings on the metal layers are symmetric for the two couplers, 
such that their perturbation to the axial \gls{rf} field is canceled at the center of the interaction zone where the beams are focused. 

The other junction trap features different optics on its long arm and short arms.
On its long arm we implemented a five-ion quantum register, shown in \fref{fig:junction_coupler}(b). 
The five couplers for \nm{729} light are designed to focus the beams to \um{2} beam waist along the trap axis,
with the beam foci matching the positions of five ions in an harmonic potential with an axial center-of-mass mode frequency of \kHz{422}~\cite{Mehta2019}.
The same coupler for \nm{854} and \nm{866} light as in \fref{fig:junction_coupler}(a) resides between the two \nm{729} couplers on the left side.
The openings on the metal layers have mirror symmetry about the trap axis.
The \gls{rf} electrodes are deformed to compensate the perturbation of pseudo-potential from the openings on them for light transmission~\cite{Mehta2019}.
On each of the three short arms, we placed one single-ion addressing optics for \nm{729} light and one coupler for repumper light the same as \fref{fig:junction_coupler}(b), as shown in \fref{fig:junction_coupler}(c).

\section{Trap chip layout and fabrication}
\label{sec:chip_layout}

We fabricated the integrated-optics traps at a commercial photonics foundry (LioniX International) using \SiN{} technology~\cite{LioniX2015}.
The layer stackup of this technology has previously been reported in \cite{Mehta2020,Mehta2019}.
The trap electrodes are patterned on \nm{300} thick top gold layer which lies on $\textrm{SiO}_2$.
A platinum layer forms an \gls{rf} ground plane \um{3} below the gold
to isolate the \SiN{} layers and silicon substrate from the \gls{rf} tracks. 
Here we also used this platinum layer to route the electric connections to the segmented middle electrodes.
The \SiN{} layers for the integrated optics are \um{3.5} below the platinum plane.

\begin{figure}[htbp]
    \centering
    \includegraphics[width=0.8\textwidth]{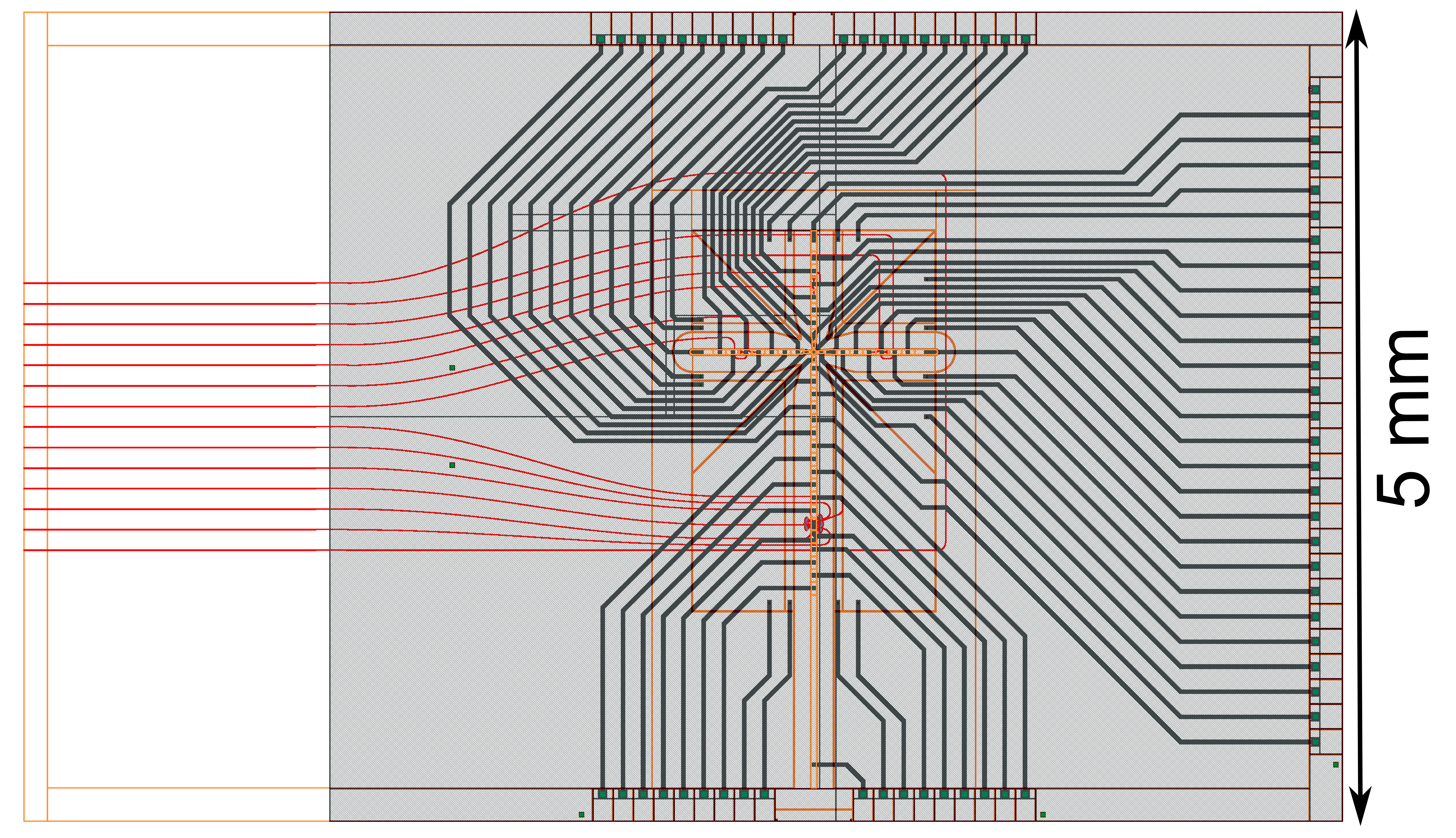}
    \caption{The layout of the junction trap chip.
    The left edge is dedicated to the fiber array for optical input.
    The waveguides (red) route the light to the interaction zones on each junction arm.
    Wirebonding pads for electrical input are patterned on the other three edges, 
    allowing for free-space beam access along $\pm\ang{45}$ on each arm and one extra axial beam along the long arm of the junction.
    The trap electrodes and the wirebonding pads are pattern on the top gold layer (outlined in gold).
    Except for the \gls{rf} drive, the electric connections from the wirebonding pads to the corresponding electrodes are routed on the platinum layer (shaded).
    The rest of the platinum layer is made a ground plane. 
    Vias (green) connect the gold and the platinum layers. 
    }
    \label{fig:junction_layout}
\end{figure}

The layout of the trap chip is created in the \textsc{Cadence Virtuoso\texttrademark{}} layout suite.
To create electrodes of irregular shapes, 
we sampled their outlines to line segments, 
exported the coordinates of the sample points to text files,
and used a script to connect them into polygons in \textsc{Cadence Virtuoso}.
The full chip layout is shown in \fref{fig:junction_layout}.

The trap chip is \mm{5} wide and approximately \mm{8.2} long.
The trap electrodes are patterned on the top gold layer, outlined in gold.
The left edge of the chip is dedicated to the fiber arrays for optical input.
The \SiN{} waveguides (red) in-couple the optical inputs from single-mode fibers attached to the edge
and route them to the interaction zones located on the four junction arms.

The wirebonding pads (\um{200}-by-\um{120}) for electric inputs are located on the other three edges of the chip.
They are patterned such that $\pm\ang{45}$ free-space beam accesses are allowed on each junction arm,
and one axial beam is allowed along the long arm.

The \gls{rf} electrodes directly extend to their wirebonding pad on the gold layer.
The electric connections from the rest of the electrodes to their wirebonding pads are routed on the platinum layer (shaded grey).
The platinum tracks are \um{10} wide, limited by the density of connections in proximity of junction center.
The clearance between the platinum tracks and the neighbouring ground plane is \um{5}.
To minimize the exposure of the \gls{rf} electrodes to the underlying silicon substrate from such \um{5} gaps, 
we tried to reduce the overlap between the platinum tracks and the \gls{rf} electrodes while routing.
To reduce the crosstalk between neighbouring tracks, 
we arranged at least \um{30} wide ground plane between neighbouring tracks.
The tracks were routed manually, and the rest of the platinum ground plane was created using the Boolean logics of \textsc{Cadence Virtuoso}.

Vias (green) form the electric connection between the gold and the platinum layers.
Square vias \um{30} wide connect the gold wirebonding pads to the platinum tracks.
We reduced the width of vias to \um{5} on the trap electrodes,
since they indent the top gold surface and can potentially cause trapping potentials to deviate from the simulations.

All blank space on the top gold layer is covered by a ground plane.
The platinum ground plane is removed over the region where the vertical tapering of the \SiN{} waveguides take place.
This is to reduce the insertion loss of input light before the optical modes are adiabatically coupled to the tightly confining thick \SiN{} waveguides. 
The two ground planes are electrically connected by five small vias \um{5} wide,
as well as a bigger one \um{30} wide on the corresponding wirebonding pad.
The wirebonding pad is used to connect the ground planes to the ground reference of the experiment, 
or alternatively to a voltage source for applying a bias voltage.

\begin{figure}[htb]
    \centering
    \includegraphics[width=0.6\textwidth]{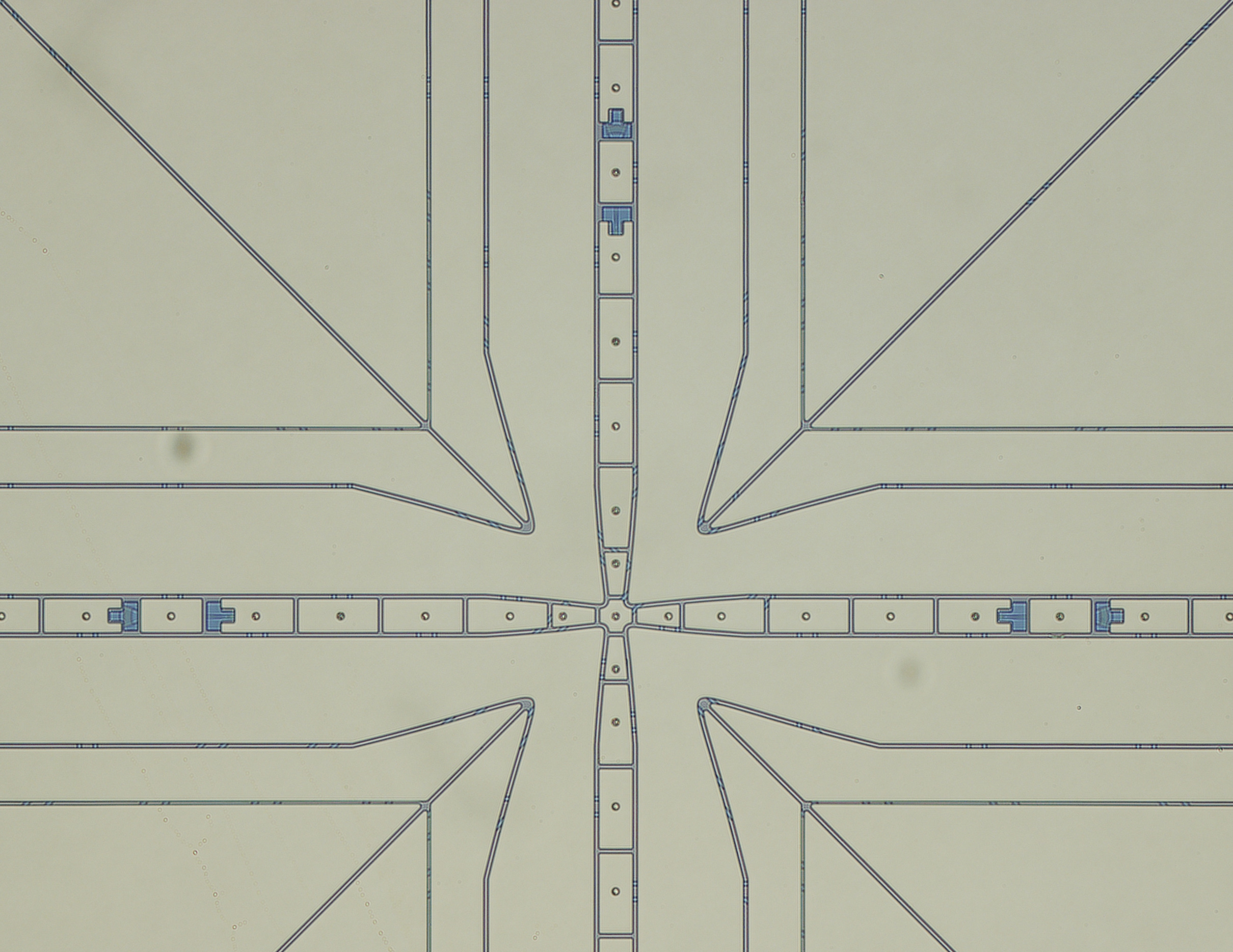}
    \caption{Optical micrograph of the fabricated junction trap.
    The vias are visible due to the indentation of the top gold surface they cause.}
    \label{fig:junction_picture}
\end{figure}

\Fref{fig:junction_picture} shows an optical micrograph of the finished junction trap featuring the integrated optics in \fref{fig:junction_coupler}(c).

\section{Summary and Outlook}
\label{sec:summary}
We presented the design of a \gls{se} junction trap based on a numerical optimization of the trap electrode shape,
parameterized using B-spline functions.
Equal weights were used in the bi-objective cost function to minimize the pseudo-potential curvature variation and the axial pseudo-potential gradient along the ion transportation path at a fixed ion height. 
The optimization was based on \textit{SurfacePattern}, and the final result was verified with an \gls{fem} simulation.
To facilitate parallel operation in multiple trap zones,
we implemented integrated optics on each arm of this X-junction.
The traps are fabricated in a commercial foundry.
The linear traps fabricated on the same wafers and the optics integrated on them have been tested experimentally in other work, serving as the basic proof of principle for the photonic foundry processes.

In an actual junction transport experiment, modifications of the transport paths and waveforms are likely needed relative to those presented in this article.
The \gls{dac} update rate~\cite{Blakestad2011,Shu2014} and slew rate~\cite{Clercq2015} can be key aspects in low-excitation junction transport,
which we have not considered in the preliminary waveform yet, 
and would be closely related to the exact experimental parameters.

Depending on the exact experimental limitations, the weights in the bi-objective optimization of the \gls{rf} electrodes can be adjusted in the future iterations of junction design,
and other objectives may be introduced as well.

To operate the current version of junction traps, the ultra-violet lasers for the cooling, detection and photo-ionization of $^{40}\textrm{Ca}^+$ still rely on the conventional free-space delivery. 
This is due to the high material absorption of \SiN{} in the low \nm{400} wavelength range~\cite{IntegratedUV2019}. 
Aluminium oxide~\cite{IntegratedUV2019,West2019} is a CMOS-compatible material with higher bandgap energies,
and is a potential choice for the full optical integration of the lasers controlling $^{40}\textrm{Ca}^+$ qubits in future iterations. 

In case the requirement for free-space laser access is relaxed by the full optical integration,
the available space for wirebonding pads roughly doubles.
This allows a larger number of control electrodes, which can provide finer control of the trapping potentials and more flexibility in parallel operations.
For instance, the outer electrodes could be segmented along the trap axis,
so that the radial tilt of different zones on the same arm can be adjusted independently.

\Glspl{tsv}~\cite{BGAtrap2015} provide a promising approach to further increased density of control electrodes and a larger-scale ion-trap array connected by multiple junctions.
Unlike the wirebonding pads that have to be located on the perimeter of the trap chip, 
the \glspl{tsv} can make use of the surface area on the backside of the trap chip to form higher number of electrical connections.
The implementation of \gls{tsv} on silicon photonics chips is successfully demonstrated in \cite{TSV_SiPh2016},
where the \glspl{tsv} are created by masked etching and copper filling after the fabrication of the integrated photonics and before the formation of metal layers.
Therefore, such a process should, in principle, be made compatible to the \SiN{} technology utilized in our integrated-optics ion traps. 

\section{Acknowledgements}
We thank Jason Amini and Curtis Volin for helpful discussions about the design principle of the junction trap,
and Frank G\"urkaynak at ETH for support with the CAD software.
We acknowledge funding from the Swiss National Science Foundation (grant no. 200020 165555), the National Centre of Competence in Research for Quantum Science and Technology (QSIT), ETH Z\"urich.

\section*{References}
\bibliographystyle{unsrt.bst}
\bibliography{reference.bib}

\end{document}